\documentclass[12pt]{article}
\usepackage[final]{graphics}
\usepackage{amsmath}
\usepackage{amsfonts,amsbsy}
\usepackage[psamsfonts]{amssymb}
\usepackage{graphicx}
\usepackage{color}

\def\empile#1\above#2{\mathrel{\mathop{\kern 0pt#1}\limits_{#2}}}

\newcommand{\non}{\nonumber\\}

\newcommand{\slp}{\raise.1ex\hbox{$/$}\kern-.63em\hbox{$p$}}
\newcommand{\slq}{\raise.1ex\hbox{$/$}\kern-.53em\hbox{$q$}}
\newcommand{\slv}{\raise.1ex\hbox{$/$}\kern-.63em\hbox{$v$}}
\newcommand{\slR}{\raise.15ex\hbox{$/$}\kern-.53em\hbox{$R$}}
\newcommand{\slQ}{\raise.15ex\hbox{$/$}\kern-.53em\hbox{$Q$}}
\newcommand{\slK}{\raise.15ex\hbox{$/$}\kern-.53em\hbox{$K$}}
\newcommand{\slk}{\raise.15ex\hbox{$/$}\kern-.53em\hbox{$k$}}
\newcommand{\slSigma}{\raise.15ex\hbox{$/$}\kern-.53em\hbox{$\Sigma$}}
\newcommand{\slcalP}{\raise.15ex\hbox{$/$}\kern-.63em\hbox{$\cal P$}}
\newcommand{\slA}{\raise.15ex\hbox{$/$}\kern-.73em\hbox{$A$}}
\newcommand{\slbfA}{\raise.15ex\hbox{$/$}\kern-.73em\hbox{${\imb A}$}}
\newcommand{\sla}{\raise.15ex\hbox{$/$}\kern-.53em\hbox{$a$}}
\newcommand{\slb}{\raise.15ex\hbox{$/$}\kern-.53em\hbox{$b$}}
\newcommand{\slc}{\raise.15ex\hbox{$/$}\kern-.53em\hbox{$c$}}
\newcommand{\slD}{\raise.15ex\hbox{$/$}\kern-.53em\hbox{$D$}}
\newcommand{\slC}{\raise.15ex\hbox{$/$}\kern-.53em\hbox{$C$}}

\def\P{{\boldsymbol P}}
\def\p{{\boldsymbol p}}
\def\q{{\boldsymbol q}}
\def\l{{\boldsymbol l}}
\def\k{{\boldsymbol k}}

\def\x{{\boldsymbol x}}
\def\y{{\boldsymbol y}}
\def\X{{\boldsymbol X}}

\def\r{{\boldsymbol r}}

\def\b{{\boldsymbol b}}
\def\a{{\boldsymbol a}}

\def\u{{\boldsymbol u}}
\def\v{{\boldsymbol v}}

\def\wt{\widetilde}
\def\bs{\boldsymbol}

\begin{document}

\thispagestyle{empty}
\title {\bf Heavy quark pair production in high\\
energy pA collisions: Quarkonium}

\author{Hirotsugu Fujii and Kazuhiro Watanabe}
\maketitle
\begin{center}
Institute of Physics, University of Tokyo,\\ 
Komaba 3-8-1, Tokyo 153-8902, Japan
\end{center}

\begin{abstract}
\noindent
Quarkonium production in high-energy proton (deuteron)-nucleus
collisions is investigated in the color glass condensate framework.
We employ the color evaporation model assuming 
that the quark pair produced from dense small-$x$
gluons in the nuclear target bounds into a quarkonium outside
the target. 
The unintegrated gluon distribution at small Bjorken $x$ 
in the nuclear target is treated with 
the Balitsky-Kovchegov equation with running coupling corrections.
For the gluons in the proton,
we examine two possible descriptions, 
unintegrated gluon distribution and ordinary collinear gluon distribution.
We present 
the transverse momentum spectrum and nuclear modification factor
for J/$\psi$ production at RHIC and LHC energies, 
and those for $\Upsilon(1S)$ at LHC energy,
and discuss the nuclear modification factor and the momentum
broadening by changing the rapidity and the initial saturation scale.  
\end{abstract}

\section{Introduction}

%
High-energy proton-nucleus (pA) 
collisions allow us to explore the dense gluon system
appearing at small values of the Bjorken $x$ in the target nucleus.
Such a dense gluon system
is expected to possess a universal feature of {\it parton saturation},
characterized by the saturation momentum scale $Q_s^2(x)$,
and has been investigated with the Color Glass Condensate 
(CGC) effective theory\cite{Gelis:2010nm,McLerV}.
In a heavy nucleus with atomic mass number $A$,
the saturation phenomenon will become relevant even at moderate $x$
because of its larger gluon density
by a factor of target thickness $A^{1/3}$. 
%
Very recently p+Pb collisions at the center-of-mass energy 
$\sqrt{s}=5.02$ TeV have been delivered
at the large hadron collider (LHC), and new exciting data are
being reported such as hadron
multiplicity\cite{ALICE:2012xs}
and ridge phenomenon\cite{CMS:2012qk,ALICE:2012ola}.
The observed hadron multiplicity and momentum spectrum
\cite{ALICE:2012xs}
will constrain theoretical models.

At this high energy, the relevant value of $x_{1,2}$ becomes so small
that the parton saturation scale $Q_s^2(x_2)\propto A^{1/3}x_2^{-\lambda}$ 
with $\lambda \approx 0.3$\cite{StastGK1,GelisPSS1}  
in the nuclear target will be
comparable to or larger than the charm quark mass~$m_c$.
This suggests coherence effects even in charm quark production,
and thus we can get useful information of 
the gluon saturation in the nuclear target
by studying the energy and rapidity dependences of charm quark and quarkonium
productions in pA collisions at relativistic heavy ion collider (RHIC)
and the LHC\cite{KharzT1,KharzT4,FujiiGV1,FujiiGV2,
Dominguez:2011cy,Kharzeev:2012py}.

In heavy ion collision experiments,
heavy quark and quarkonium productions\cite{Bedjia1,Brambilla:2004wf}
are very valuable probes for quantifying properties of hot and dense
matter or a quark-gluon plasma transiently created in the events.
Quarkonium suppression\cite{MatsuS1} and 
enhancement\cite{Thews:2000rj,BraunMunzinger:2000ep}, 
and energy loss\cite{Dokshitzer:2001zm,Djordjevic:2003zk,Armesto:2005mz} 
and collective flow of heavy flavor mesons\cite{vanHees:2004gq}
have already been discussed extensively.
Here one obviously needs to know the initial nuclear effects on their
productions for quantitative understanding of hot-medium modifications.

%
In this paper we shall investigate phenomenological implications
of the saturation effects on the quarkonium production at collider
energies by exploiting the color evaporation model 
(CEM)\cite{Brambilla:2004wf}.  
Systematic study of quarkonium production spectrum from CGC in pA collisions
will quantitatively improve our understanding of gluon saturation
effects on particle production in pA collisions. 
At the same time, it will serve as a benchmark for assessing the hot
medium effects on the quarkonium production in AA collisions.

%
In the CGC framework, the quark-pair production cross section
in pA collisions is obtained by Blaizot, Gelis and 
Venugopalan~\cite{BlaizGV1,BlaizGV2},
where pA is treated as a ``dilute-dense'' system and 
the cross section is evaluated 
at leading order in the coupling constant $\alpha_s=g^2/4 \pi$ and 
the color charge density $\rho_p$ in the proton, but
in full orders with respect to the color charge density 
$g^2 \rho_A=O(1)$
in the nucleus.
Adopting this formula, we previously evaluated
the heavy quark production cross-sections in high-energy pA
collisions to reveal their general features\cite{FujiiGV1,FujiiGV2}.

%
Of particular importance is 
the $x$ dependence of the unintegrated gluon distribution (uGD) in
the nuclear target. 
We describe the $x$ dependence of uGD 
with the Balitsky-Kovchegov (BK) equation including the running
coupling corrections (rcBK equation)\cite{AlbacK1}.
The BK equation can be derived from a more general equation of
functional renormalization group by taking 
the mean-field approximation.
The authors of \cite{Albacete:2009fh,Albacete:2010sy}
analyzed the global data of deep inelastic scatterings (DIS) of e+p 
at HERA using the rcBK equation,
to obtain a set of constrained uGD of the proton at small $x<0.01$.
The resultant uGD has been applied to compute the
particle production at the proton-(anti)proton
colliders\cite{ALbacete:2010ad,Albacete:2012xq}.
We use this constrained uGD in this work.

%
In forward particle production,
the momentum fraction $x_1$ of the gluons from the proton is not small,
and the distribution may be better described with the ordinary 
collinear gluon distribution function.
Accordingly, the expression of the cross section is obtained by 
taking the collinear limit $\k_{1\perp} \to 0$ 
for the gluon momentum from the proton in the hard matrix element.
(This kind of asymmetric treatment is well known for the hadron production 
from CGC\cite{DumitHJ1}.)
We will compare the quarkonium production in
the collinear approximation on the
proton side with the original one that involves
the $\k_{1\perp}$ dependent uGD for the proton.

This paper is organized as follows: In Sec.~2 we review the quark pair
production formula in the CGC framework
and explain its collinear limit on the proton side.
Then we incorporate the $x$-dependence of uGD 
using the rcBK evolution equation.
In Sec.~3, we compute the quarkonium production cross section
working in the color evaporation model, and 
show the transverse momentum spectra and 
nuclear modification factor for J/$\psi$
at RHIC and LHC energies, and those for $\Upsilon$
at LHC energy.
Rapidity and initial-saturation-scale dependences
of the nuclear modification factor, and momentum
broadening are also discussed.
Sec.~4 is devoted to conclusion and outlook.

\section{Pair production cross-section}
In this section we summarize analytic expressions for the quark-pair
production cross-section, derived in Ref.~\cite{BlaizGV2}. We also
present its limit when the transverse momentum of the gluon coming
from the proton 
is small, in order to make use of conventional collinear factorization
for the proton.  For more details, see Refs.~\cite{BlaizGV2,FujiiGV2}.

\subsection{Pair cross-section in the large $N$ limit}
In the CGC formalism, the proton-nucleus collision is described as a
collision of two sets of color sources representing the large $x$
degrees of freedom in the proton and the nucleus respectively. When
they collide, these color sources produce a time-dependent classical
color field, and this color field can in turn produce quark-antiquark
pairs. Both the classical color field and the quark-pair production
amplitude can be calculated analytically when one of the projectiles
is dilute and its density of color sources can be treated to the lowest
order. This is the approximation we make in the description of
proton-nucleus collisions in this framework, the proton naturally
being the dilute projectile. 

For definiteness, let us denote by $\rho_p$ and $\rho_{_A}$ the densities
of color sources in the proton and the nucleus respectively, and that
the proton moves in the $+z$ direction and the nucleus in the $-z$
direction. Then, the pair production amplitude reads \cite{BlaizGV2}~:
\begin{eqnarray}
&&\!\!\!\!\!\!\!\!\!\!\!\!
{\cal M}_{_{F}}(\q,\p)\!=\!g^2\!\int\!\frac{d^2\k_{1\perp}}{(2\pi)^2}
\frac{d^2\k_\perp}{(2\pi)^2}
\frac{\rho_{p,a}(\k_{1\perp})}{k_{1\perp}^2}
\!\int\!\! d^2\x_\perp d^2\y_\perp
e^{i\k_\perp\cdot\x_\perp}
e^{i(\p_\perp\!+\!\q_\perp\!-\!\k_\perp\!-\!\k_{1\perp})\cdot\y_\perp}
\non
&&\quad\times
\overline{u}(\q)\Bigg\{ T_{q\bar{q}}(\k_{1\perp},\k_{\perp})
[{\wt U}(\x_\perp)t^a {\wt U}^\dagger(\y_\perp)]
+T_{g}(\k_{1\perp})[t^bU^{ba}(\x_\perp)]\Bigg\} v(\p)
\; ,
\label{eq:Mf-final-1}
\end{eqnarray}
where we denote\footnote{The momenta $p$ and $q$ of the produced
  particles have not been listed among the arguments of these objects
  in order to make the equations more compact.}
\begin{eqnarray}
&&T_{q\bar{q}}(\k_{1\perp},\k_{\perp})\equiv
\frac{\gamma^+(\slq-\slk+m)\gamma^-(\slq-\slk-\slk_1+m)\gamma^+}
{2p^+[(\q_\perp\!-\!\k_\perp)^2+m^2]+2q^+[(\q_\perp\!-\!\k_\perp\!-\!\k_{1\perp})^2+m^2]}
\; , 
\non
&&T_{g}(\k_{1\perp})\equiv
\frac{\slC_{_{L}}(p+q,\k_{1\perp})}{(p+q)^2}
\; ,
\label{eq:Tqqbar-Tg}
\end{eqnarray}
with $\k_{2\perp} \equiv \p_\perp + \q_\perp - \k_{1\perp}$, and where
$C_L^\mu(p+q,\k_{1\perp})$ is the well-known Lipatov effective
vertex\footnote{Its components are~:
\begin{equation*}
C_{_L}^+(q,\k_{1\perp}) \equiv \frac{-k_{1\perp}^2}{q^-}+ q^+\,\,;\,\,
C_{_L}^-(q,\k_{1\perp}) \equiv \frac{k_{2\perp}^2}{q^+}-q^-\,\,;\,\,
C_{_L}^i(q,\k_{1\perp}) \equiv -2k_1^i +q^i \; .
\label{eq:Lipatov}
\end{equation*}}.
The matrix ${\wt U}(\x_\perp)$ is the path-ordered exponential of the
gauge field generated by the charge density $\rho_{_A}$ in the nucleus:
\begin{eqnarray}
{\wt U}(\x_\perp)\equiv {\cal P}_+ \exp\left[-ig^2\int_{-\infty}^{+\infty}
dz^+ \frac{1}{{\bs\nabla}_{\perp}^2}\,\rho_{_A}(z^+,\x_\perp)\cdot t
\right] \, ,
\end{eqnarray}
with $t^a$ the SU($N$) generators in the fundamental representation,
and $U(\x_\perp)$ is the same but in the adjoint representation.  The
two terms in the curly bracket of Eq.~(\ref{eq:Mf-final-1}) may be
interpreted as the processes where a gluon from the proton emits the
pair either before or after the collision with the nucleus.  The
eikonal phases ${\wt U}$ and $U$ describe the multiple scatterings of
a quark or a gluon in the target nucleus.  An important property of
this amplitude is that the sum of the two terms in the bracket
vanishes when $\k_{1\perp}\to 0$ in agreement with Ward identities
\cite{BlaizGV2} -- this property is essential in order to recover the
limit of collinear factorization on the proton side, which should hold
since there is only a single scattering on the proton.

The probability of pair production is then obtained by squaring the
above amplitude, and by averaging the squared amplitude with the
classical charge distributions of the proton and the nucleus, $W_p[x,
\rho_p]$ and $W_{_A}[x,\rho_{_A}]$.  The resulting expression
generally involves the nuclear multi-parton correlators, e.g., $
\langle {\rm tr}\big ( {\wt U}(\x_\perp)t^a {\wt U}^\dagger(\y_\perp)
{\wt U}(\y^\prime_\perp)t^a {\wt U}^\dagger(\x^\prime_\perp) \big )
\rangle_{_Y}$, where $\langle \cdots \rangle_{_Y}$ indicates that the
average over $\rho_{_A}$ is performed with the distribution
$W_{_A}[x,\rho_{_A}]$ evolved to the rapidity $Y\equiv\ln(1/x)$.  
2- and 3-point correlators are also needed in the calculation of the
pair production probability, but they can be obtained as special cases
of the 4-point function thanks to the identity ${\wt
U}(\x_\perp)t^a{\wt U}^\dagger(\x_\perp) =t^b U^{ba}(\x_\perp)$.  This
formula also leads to a general sum rule for the Fourier transforms of
these correlators $\phi_{_{A,Y}}^{q\bar{q},q\bar{q}}$,
$\phi_{_{A,Y}}^{q\bar{q},g}$ and $\phi_{_{A,Y}}^{g,g}$
\cite{BlaizGV2,FujiiGV2}.  

All these correlators can be evaluated in a closed form when the
distribution of color sources, $W_{_A}[x,\rho_{_A}]$, is a
Gaussian\footnote{This distribution is a Gaussian in the
McLerran-Venugopalan model for a large nucleus, and also in the
asymptotically small $x$ regime at very high energy.  We therefore
expect that this is a reasonable approximation for our work.}  (see
\cite{BlaizGV2}), but the 4-point correlator has a very complicated
expression which is quite hard to evaluate numerically. However, in
the large $N$ limit, it simplifies into
\begin{eqnarray}
{\rm tr}\big<{\wt U}(\x_\perp)t^a {\wt U}^\dagger(\y_\perp)
             {\wt U}(\u_\perp)t^a {\wt U}^\dagger(\v_\perp)  \big>_{_Y}  
&\!\!\!\!\! 
\empile{=}\above{N\to \infty}
&\!\!\!\!\!
\frac{N^2}{2}
S_{_Y}(\x_\perp, \v_\perp)
S_{_Y}(\u_\perp, \y_\perp)
\label{eq:largeN}
\end{eqnarray}
with
\begin{eqnarray}
S_{_Y}(\x_\perp,\y_\perp)\equiv
{1 \over N}{\rm tr}\big<{\wt U}(\x_\perp) 
{\wt U}^\dagger(\y_\perp)\big>_{_Y}  \; .
\end{eqnarray}
As one can see, it is possible in this limit to write the 4-point (and
also the 3-point) function in terms of the 2-point function only,
which simplifies considerably all the numerical calculations.
Together with the translational invariance in the transverse plane,
this fact makes the relation between
$\phi_{_{A,Y}}^{q\bar{q},q\bar{q}}$ and $\phi_{_{A,Y}}^{q\bar{q},g}$
trivial.

By exploiting these relations between the 4-, 3- and 2-point
correlators, the pair production probability at the impact parameter
$\b$, in the large $N$ limit, can be written as~\cite{FujiiGV2}
\begin{eqnarray}
&&
\frac{d P_1(\b)}{d^2\p_\perp d^2\q_\perp dy_p dy_q}
=
\frac{\alpha_s^2 N}{8\pi^4 d_A}
\frac{1}{(2\pi)^2}
\non
&&\qquad\times
\int\limits_{\k_{2\perp},\k_\perp}\!\!
\frac{\Xi(\k_{1\perp},\k_{2\perp},\k_{\perp})}
{\k_{1\perp}^2 \k_{2\perp}^2}
\;
\frac{d\phi_{_A,y_2}^{q\bar{q},g}(\k_{2\perp},\k_\perp;\b)}
     {d^2 \X_\perp}
\; 
\varphi_{p,y_1}(\k_{1\perp}) \;  ,
\label{eq:average-number-LN}
\end{eqnarray}
where we denote $\int_{\k_\perp} \equiv \int d^2 \k_\perp / (2\pi)^2$,
$d_{A}\equiv N^2-1$ the dimension of the adjoint representation of
$SU(N)$, and $\k_{1\perp}=\p_\perp+\q_\perp - \k_{2\perp}$.  The
variables $y_{1,2}$ are the rapidities of the gluons that come from
the proton and from the nucleus respectively. A shorthand notation for
the squared matrix element is introduced as\footnote{In general, the
Fourier transform of the 4-point function depends on three momentum
variables: $\k_{2\perp}$, $\k_\perp$ and $\k_\perp^\prime$ (see
\cite{BlaizGV2}). However, in the large $N$ limit, this 4-point
function is in fact given by (see \cite{FujiiGV2})
\begin{equation*}
\frac{d\phi_{_A,y_2}^{q\bar{q},q\bar{q}}(\k_{2\perp},\k_\perp,\k_\perp^\prime;\b)}{d^2
\X_\perp}= (2\pi)^2\delta(\k_\perp-\k_\perp^\prime)
\frac{d\phi_{_A,y_2}^{q\bar{q},g}(\k_{2\perp},\k_\perp;\b)}{d^2
\X_\perp}\; ,
\end{equation*} which allowed us to equate $\k_\perp$ and $\k_\perp^\prime$ in the squared amplitude and to perform directly 
the $\k_\perp^\prime$ integration.}
\begin{eqnarray}
\Xi(\k_{1\perp},\k_{2\perp},\k_{\perp})
&=&
{\rm tr}_{\rm d}
\Big[(\slq\!+\!m)T_{q\bar{q}}(\slp\!-\!m)
\gamma^0 T_{q\bar{q}}^{\dagger}\gamma^0\Big]
\non
&&
+
{\rm tr}_{\rm d}
\Big[(\slq\!+\!m)T_{q\bar{q}}(\slp\!-\!m)
\gamma^0 T_{g}^{\dagger}\gamma^0 + {\rm h.c.}\Big]
\non
&&
+{\rm tr}_{\rm d} 
\Big[(\slq\!+\!m)T_{g}(\slp\!-\!m)\gamma^0 T_{g}^{\dagger}\gamma^0\Big]
\; .
\label{eq:Xi}
\end{eqnarray}
In Eq.~(\ref{eq:average-number-LN}), $\varphi_{p,y_1}$ is the
uGD in the proton, and
${d\phi_{_A,y_2}^{q\bar{q},g}}/{d^2 \X_\perp}$ is expressed in terms
of the Fourier transform of the 3-point nuclear correlator
as\footnote{In Eq.~(\ref{eq:average-number-LN}), this object appears
in differential form with respect to the transverse coordinate
$\X_\perp$ because we are considering here the pair production
probability at a fixed impact parameter $\b$. When we integrate it
over $\b$ in order to obtain the cross-section, we will have to
integrate this correlator over the transverse area of the nucleus.}~:
\begin{eqnarray}
\frac{d \phi_{_A,_Y}^{q \bar{q},g}(\l_\perp,\k_\perp;\X_\perp)}{d^2 \X_\perp}
&\equiv&
\frac{\l^2_\perp }{2 N \alpha_s} \; 
\int\limits_{\x_\perp,\y_\perp}
e^{i\k_\perp \cdot (\x_\perp -\x^\prime_\perp)}
e^{i(\l_\perp-\k_\perp)\cdot (\y_\perp-\x^\prime_\perp)}
\non
&&\quad\times
\langle {\rm tr}\big (
 \tilde U(\x_\perp)t^a \tilde U^\dagger(\y_\perp)
 \tilde U(\x^\prime_\perp)t^a \tilde U^\dagger(\x^\prime_\perp) \big )
\rangle_{_Y}
\non
&\empile{=}\above{{\rm large}~N}&
\frac{N\l^2_\perp}{4 \alpha_s} \; 
 S_{_Y}(\k_\perp) \;
S_{_Y}(\l_\perp-\k_\perp)\;  , 
\label{eq:3ptfn}
\end{eqnarray}
where $S_{_Y}(\k_\perp)$ is the Fourier transform of
$S_{_Y}(\x_\perp)$.  The $\X_\perp$ dependence is rather weak for a
large nucleus and may be treated implicitly through the variations of
saturation scale $Q_{s,A}^2(\X_\perp)$ with $\X_\perp$.  In the second
line of Eq.~(\ref{eq:3ptfn}), we have ignored the
$\X_\perp$-dependence of $\phi_{A,y}^{q\bar{q},g}$ when doing the
$\X_\perp$ integration because the proton radius is small compared
with that of a heavy nucleus ($R_p \ll R_{_A}$). We thus have a
compact expression for the quark production probability, but our
formula involves a nuclear 3-point function
$\phi_{_A,y}^{q\bar{q},g}$, which violates the usual form for
$k_\perp$-factorization even in the leading order
approximation\cite{BlaizGV2,FujiiGV1}.

The pair cross-section in the minimum-bias pA collision is obtained by
integrating Eq.~(\ref{eq:average-number-LN}) over the impact parameter
$\b$. Dividing the cross-section with the total inelastic
cross-section $\sigma_{hadr}^{\rm pA}$, which we estimate as
$\sigma_{hadr}^{\rm pA} = \pi (R_A + R_p)^2 \approx\pi R_A^2$, we have the
average multiplicity per event~:
\begin{eqnarray}
\frac{d N_{q \bar{q}}}{d^2\p_\perp d^2\q_\perp dy_p dy_q}
&\!\!\!\!\!=&\!\!\!\!\!
\frac{1}{\pi R_A^2}
\frac{\alpha_s^2 N}{8\pi^4 d_A}
\frac{1}{(2\pi)^2}
\int\limits_{\k_{2\perp},\k_\perp}\!\!\!
\frac{\Xi(\k_{1\perp},\k_{2\perp},\k_{\perp})}
{\k_{1\perp}^2 \k_{2\perp}^2}
\;
\phi_{_A,y_2}^{q\bar{q},g}(\k_{2\perp},\k_\perp)
\; 
\varphi_{p,y_1}(\k_{1\perp})
\; .
\non
\label{eq:cross-section-LN}
\end{eqnarray}
Here we have introduced the 3-point function integrated
over nuclear transverse area
\begin{eqnarray}
\phi_{_A,_Y}^{q \bar{q},g}(\l_\perp,\k_\perp)
&=&
\pi R_A^2 \; \frac{N\l^2_\perp}{4 \alpha_s} \; 
 S_{_Y}(\k_\perp) \;
S_{_Y}(\l_\perp-\k_\perp)\; ,  
\label{eq:3ptfn_2}
\end{eqnarray}
which is related to the uGD (2-point function) of the nucleus by
$\phi_{_A,_Y}^{g,g}(\l_\perp)
=\int_{\k_\perp}  \phi_{_A,_Y}^{q \bar{q},g}(\l_\perp,\k_\perp)$.
The proton uGD $\varphi_{p,y}$ 
may be estimated by replacing the transverse area
$\pi R_A^2$ and the amplitude $S_Y$ with those for the proton.

\subsection{Collinear limit on the proton side}   %
When the momentum fraction $x_1$ probed in the proton is not small
(e.g., $x_1>10^{-3}$), and even more so in the forward rapidity region
where $x_1 = {\cal O}(1)$, the typical transverse momentum of the
gluons in the proton is much smaller than the transverse mass of the
produced quark or antiquark, $m_{\p \perp}  \gg k_{1\perp}
={\cal O}(\Lambda_{_{\rm QCD}})$.  We can therefore neglect
$\k_{1\perp}$ in the matrix element $\Xi$ in Eq.~(\ref{eq:Xi}), and
take the collinear approximation on the proton side.  This limit is
well defined thanks to the fact that the expression on the second line
in the amplitude in Eq.~(\ref{eq:Mf-final-1}) goes to zero as
$\k_{1\perp}\to 0$ \cite{BlaizGV2}~:
\begin{equation}
{\cal M}_{_F}(\q,\p)\empile{=}\above{\k_{1\perp}\to 0}
{\bs A}\cdot\k_{1\perp}+{\cal O}(\k_{1\perp}^2)\; .
\end{equation}
Thus, the amplitude squared $\Xi$ is quadratic in $\k_{1\perp}$ when
$\k_{1\perp}\to 0$, which cancels the factor $k_{1\perp}^2$ in the
denominator of Eq.~(\ref{eq:cross-section-LN}). Note that the
``vector'' ${\bs A}$ in this formula contains spinors and Dirac
matrices. In this approximation, we can write 
the integral in Eq.~(\ref{eq:cross-section-LN}) as
\begin{eqnarray}
\int\limits_{\k_{1\perp},\k_\perp}\!\!\!
\frac{{\rm tr}_{\rm d}({\bs A}^i{\bs A}^j)\; \k_{1\perp}^i \k_{1\perp}^j}
{\k_{1\perp}^2 \k_{2\perp}^2}
\;
\phi_{_A,y_2}^{q\bar{q},g}(\k_{2\perp},\k_\perp)
\; 
\varphi_{p,y_1}(\k_{1\perp})
\; ,
\non
\label{eq:cross-section-LN1}
\end{eqnarray}
where it is now implicit that $k_{1\perp}$ should not exceed the
typical transverse momentum scale set by the produced final
state. Using $d^2\k_{1\perp}=\frac{1}{2}d\theta_1 d(k_{1\perp}^2)$ and
\begin{eqnarray}
{1 \over {4\pi^3} }
\int^{Q^2} d (k_{\perp}^2) \varphi_{p,y}(\k_{\perp})
\equiv
xG_p(x=e^{-y},Q^2)\; ,
\label{eq:xG-cgc}
\end{eqnarray}
we obtain
\begin{eqnarray}
\frac{dN_{q\bar{q}}}{d^2\p_\perp d^2\q_\perp dy_p dy_q}
&\!\!\!\!\!=&\!\!\!\!\!
\frac{1}{\pi R_A^2} \;
\frac{\alpha_s^2 N}{8\pi^2 d_A}
\frac{1}{(2\pi)^2}
\!\!\int\limits_{\k_\perp}\!
\frac{\Xi_{\rm coll}(\k_{2\perp},\k_{\perp})}{\k_{2\perp}^2}
\; 
\phi_{_A,y_2}^{q\bar{q},g}( \k_{2\perp},\k_\perp)
\;
 x_1 G_p(x_1,Q^2) ,
\non
\label{eq:cross-section-LN-coll}
\end{eqnarray}
where we have now $\k_{2\perp}=\p_\perp + \q_\perp$, and where we
denote 
$\Xi_{\rm coll}(\k_{2\perp},\k_{\perp}) \equiv 
\frac{1}{2}{\rm tr}_{\rm d}({\bs A}^2)$. 
The squared matrix element 
$\Xi_{\rm coll}$ in the collinear approximation
can be obtained by expanding the amplitude in Eq.~(\ref{eq:Mf-final-1})
to linear order in the transverse momentum $\k_{1\perp}$: 
\begin{equation}
\Xi_{\rm coll}
=
\Xi_{\rm coll}^{q\bar{q},q\bar{q}}
+
\Xi_{\rm coll}^{q\bar{q},g}
+
\Xi_{\rm coll}^{g,g}\; ,
\end{equation}
with
\begin{eqnarray}
\Xi_{\rm coll}^{q\bar{q},q\bar{q}}
&=&
\frac{8p^+q^+}{(p^++q^+)^2(\a_\perp^2+m^2)^2}
\left[
m^2+\frac{(p^+)^2+(q^+)^2}{(p^++q^+)^2}\a_\perp^2
\right]\; ,
\nonumber\\
\Xi_{\rm coll}^{q\bar{q},g}
&=&
-\frac{16}{(p+q)^2(\a_\perp^2+m^2)}
\left[
m^2+\frac{(p^+)^2+(q^+)^2}{(p^++q^+)^3}\a_\perp\cdot(p^+\q_\perp-q^+\p_\perp)
\right]\; ,
\nonumber\\
\Xi_{\rm coll}^{g,g}
&=&
\frac{8}{(p+q)^4}
\left[
(p+q)^2-\frac{2}{(p^++q^+)^2}(p^+\q_\perp-q^+\p_\perp)^2
\right]\; .
\end{eqnarray}
In these formulas, we denote $\a_\perp\equiv \q_\perp-\k_\perp$, and
the squared invariant mass of the pair, $(p+q)^2$, is given by
\begin{equation}
(p+q)^2=
(p^++q^+)\left[
\frac{\p_\perp^2+m^2}{p^+}+\frac{\q_\perp^2+m^2}{q^+}
\right]-(\p_\perp+\q_\perp)^2\; .
\end{equation}

\subsection{Correlators and energy evolution}
The dense nuclear distribution of gluons is encoded in the 3-point
correlator $\phi_{_A,_Y}^{q\bar{q},g}$, that appears in the pair
cross-section (\ref{eq:cross-section-LN}) or
(\ref{eq:cross-section-LN-coll}).  In the large $N$ limit, it can be
expressed in terms of the 2-point correlation $S_{_Y}(\k_\perp)$,
i.e. the forward scattering amplitude of a dipole in the fundamental
representation.

In the quasi-classical MV model, $S_{_Y}(\k)$ is independent of the
rapidity variable $Y$, and it only includes the effects of the
multiple scatterings of the quark-antiquark pair passing though the
target nucleus, in the eikonal approximation.  At large 
$\k_\perp \gg Q_s$, 
the effect of multiple scatterings becomes small and leading
twist results are recovered.

The energy (rapidity) dependence of $S_{_Y}(\k)$ arises via quantum
fluctuations. Generally, the evolution equation for the 2-point
function involves an infinite hierarchy of multi-point correlators.
However, it is known that in the limit of a large number of colors $N$
and of a large nucleus the energy evolution of $S_{_Y}(\k_\perp)$ can
be described by a closed mean-field equation known as the
Balitsky-Kovchegov (BK) equation\cite{Balit1,Kovch1}. This equation is
an integro-differential equation that reads\footnote{We have written
it here in the approximation where the nucleus is translation
invariant in the transverse plane. This is a reasonable approximation
for a large nucleus, since the edges of the nucleus -- where this
invariance is broken -- give a comparatively small contribution to the
cross-section.}
\begin{eqnarray}
&&-\frac{d}{dY}
S_{_Y}({\r_\perp})
 =
\int d\r_{\perp 1} \,
\mathcal{K}(\r_\perp, \r_{1\perp}) \, 
\Big [
S_{_Y}({\r_\perp})
-
S_{_Y}({\r_{1\perp}})S_{_Y}({\r_{2\perp}})
\Big ]\; ,
\label{eq:Kovchegov}
\end{eqnarray}
where $\r_\perp = \r_{1\perp}+\r_{2\perp}$ and 
${\mathcal K}(\r_\perp,\r_{1 \perp})$ is the evolution kernel (see below).
Thus, with an appropriate initial condition at a certain $x=x_0$, we
can consistently treat the rapidity dependence of the cross-section by
substituting into Eq.~(\ref{eq:3ptfn})
the solution $S_{_Y}(\k_\perp)$ of the BK equation for $x<x_0$.  

It is also well known that the BK equation with a fixed coupling constant
requires a very low value of $\alpha_s$ in order
for the evolution of the saturation scale to be compatible with what
one infers from HERA data,
{\it i.e.}, $Q_s^2(Y)\sim \exp(\lambda Y)$ with $\lambda\approx 0.3$
\cite{StastGK1,GelisPSS1}. 
It was argued that the next-to-leading order
corrections to the BK equation would give the correct evolution speed
with more reasonable values of $\alpha_s$ \cite{Trian1}.
Indeed, it has been demonstrated recently 
in Refs.~\cite{AlbacK1,Albacete1} that 
the BK equation including the running coupling corrections
in the kernel in Balitsky's prescription\cite{Balit3}:
\begin{align}
\mathcal{K}(\r_\perp,\r_{1\perp})=&
\frac{\alpha_s (r^2) N} {2\pi^2}\,
\left [
\frac{1}{r_1^2} \left ( \frac{\alpha_s(r_1^2)}{\alpha_s(r_2^2)}-1  \right )
+
\frac{r^2}{r_1^2 r_2^2}
+
\frac{1}{r_2^2} \left ( \frac{\alpha_s(r_2^2)}{\alpha_s(r_1^2)}-1  \right )
\right ]
\; ,
\end{align}
makes the saturation scale behave compatible with HERA data, and
the $x$-evolution equation becomes now a very useful tool 
(called rcBK equation) for phenomenology.

Global fitting of the compiled HERA e+p data at $x<x_0=0.01$
was performed with the rcBK equation 
in \cite{Albacete:2009fh,Albacete:2010sy}. 
Following their approach, we choose the initial condition 
at $Y_0\equiv\ln(1/x_0)$ as
\begin{align}
S_{_{Y0}}(\r_\perp)=\exp \left [
-\frac{(r^2 Q_{s0,{\rm p}}^2)^\gamma }{4} \ln 
\left ( \frac{1}{\Lambda r} + e \right ) 
\right ] 
\; ,
\label{eq:initialcondition}
\end{align}
and the parameter values are listed in 
Table~\ref{tab:par}\cite{Albacete:2012xq}.
As for the running coupling constant in the evolution
kernel we adopt the following form in the coordinate space:
\begin{align}
\alpha_s(r^2)= \left [b_0 \ln 
\left (\frac{4 C^2}{r^2 \Lambda^2}+a \right ) \right ]^{-1}
\end{align}
with $b_0=9/(4\pi)$.
A constant $a$ is introduced 
so as to freeze the coupling constant smoothly at
$\alpha_s(\infty)=\alpha_{fr}$.
The non-Gaussian value $\gamma>1$ is preferred by the fitting.
It is argued in \cite{Dumitru:2011ax} that
the value $\gamma>1$ suggests a possible importance of 
higher-order color correlations in the proton
and is valid for moderate values of transverse momentum $k_{\perp}$.
We also list the McLerran-Venugopalan model $\gamma=1$ for comparison.

\begin{table}[t]
\renewcommand\arraystretch{1.2}
\begin{center}
\begin{tabular}{|c|cccc|}
\hline
set & $Q_{s0,\rm p}^2/{\rm GeV}^2$ & $\gamma$ & $\alpha_{fr}$ & $C$ \\
\hline 
g1118  & 0.1597 & 1.118 & 1.0 & 2.47 \\
MV     & 0.2    & 1     & 0.5 & 1\\
\hline
\end{tabular}
\caption{Parameter values for the two-point correlator $S_Y$.
$\Lambda=0.241$ GeV is fixed.
\label{tab:par}}
\end{center}
\end{table}

For $x_0 < x \le 1$, we extrapolate the function 
$\phi_{_A,_Y}^{q\bar{q},g}$
with the following phenomenological Ansatz~\cite{GelisSV1}:
\begin{equation}
\phi_{_A,_Y}^{q\bar{q},g}(\l_\perp,\k_\perp)=\phi_{_A,_{Y_0}}^{q\bar{q},g}(\l_\perp,\k_\perp)
\left(\frac{1-x}{1-x_0}\right)^4 \left(\frac{x_0}{x}\right)^{0.15}\; .
\end{equation}
In this formula, the power $4$ for the factor $1-x$ comes from
the behavior at large $x$ of the gluon
distributions, as inferred from sum rules. Note that this
extrapolation implies that the saturation scale is frozen at large
$x$, which may lead to a harder $\k_\perp$-spectrum for $x>x_0$ than
expected, possibly overestimating the Cronin peak.

The saturation scale for a heavy nucleus at moderate values of $x$
will be enhanced by a factor of the nuclear thickness $T_A(\b)$ :
$Q_{s,A}^2(x,\b) \propto T_A(\b)\; Q_{s,p}^2(x)$ \cite{McLerV}. 
As we consider only mean bias events in this work, 
we assume a simpler relation
\begin{eqnarray}
Q_{s,A}^2(x_0) = A^{1/3} \, Q_{s,p}^2(x_0) \; , 
\label{eq:GBW2}
\end{eqnarray}
where $A$ is the mass number of the nucleus under consideration. This
relation is assumed to be valid at the $x_0$ where the initial
condition for the rcBK equation is set. The rcBK equation then controls
how the saturation scale of the nucleus evolves to lower values of~$x$. 
Taking into account the uncertainty of the saturation scale
for a heavy nucleus $A \sim 200$ at $x=x_0$, 
we will consider the values in the range 
$Q_{s,A}^2= (4 - 6) \times Q_{s, p}^2$
at $x_0=0.01$.

\begin{figure}[tb]
\begin{center}
\resizebox*{!}{5.5cm}{\includegraphics[angle=-90]{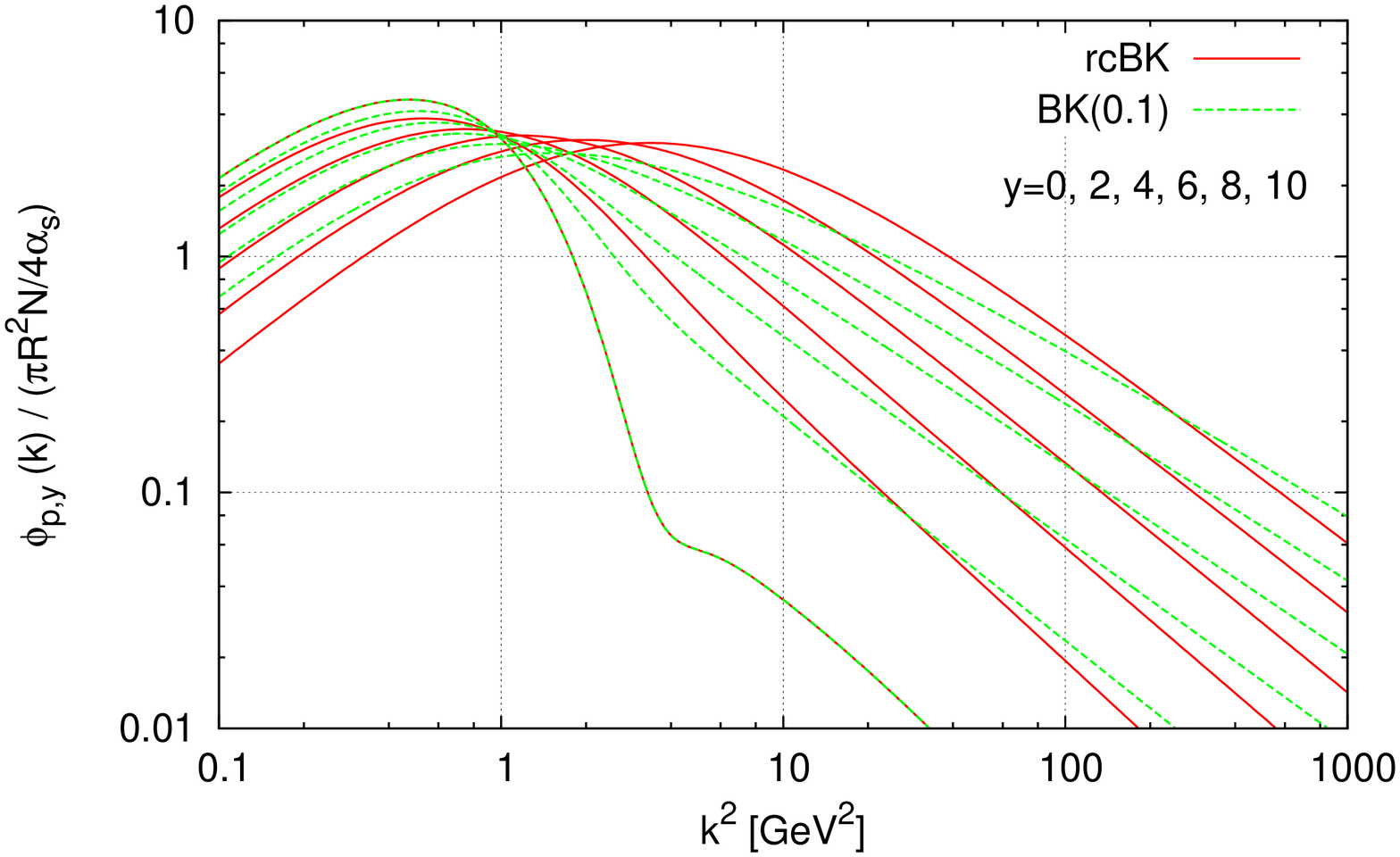}}
\resizebox*{!}{5.5cm}{\includegraphics[angle=-90]{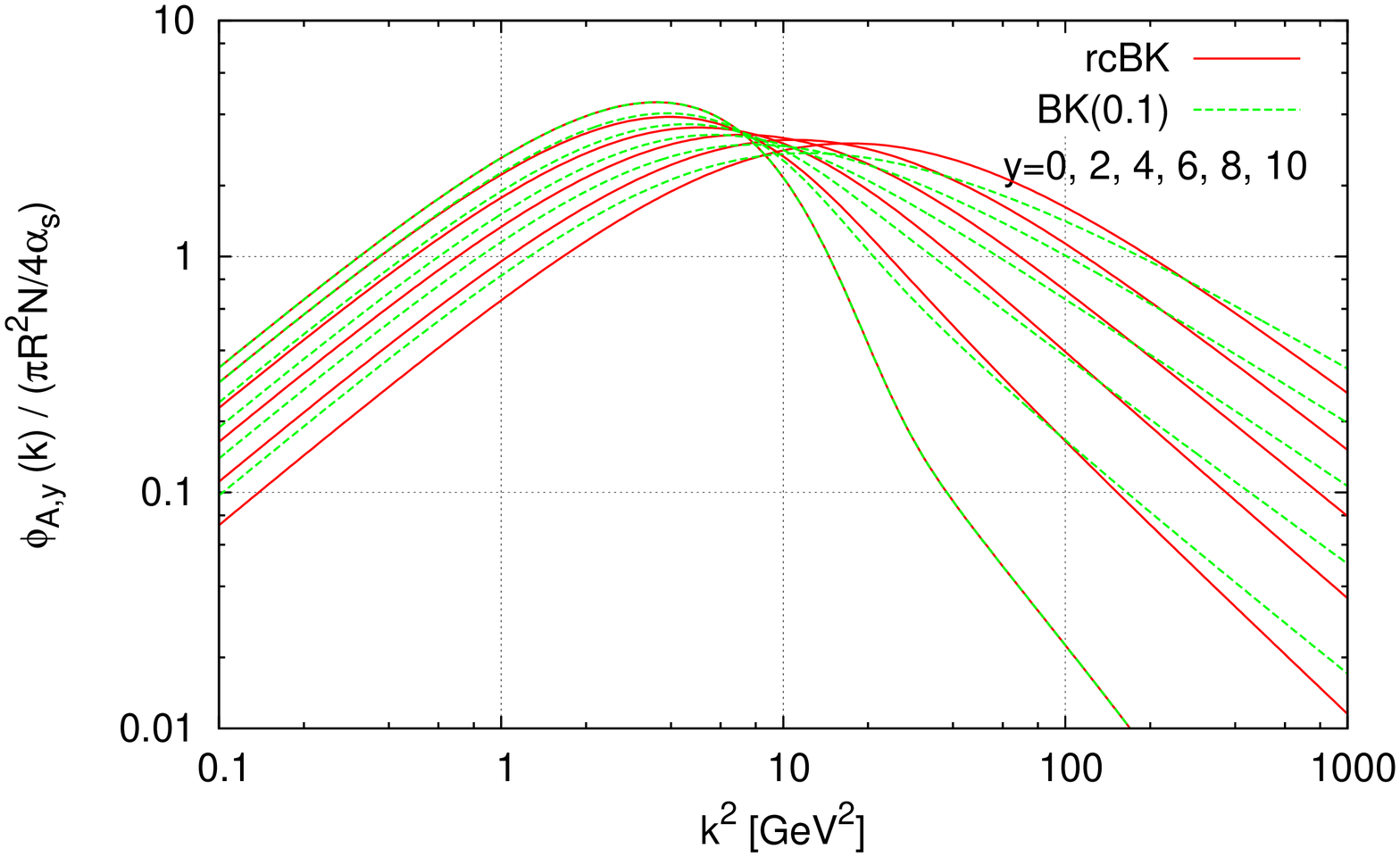}}
\end{center}
\caption{\label{fig:updf} Left:
Evolution of the proton uGD, 
$\phi_y(k_\perp^2)/(\pi R^2 N/4 \alpha_s)$ obtained
by solving the rcBK equation (solid)
and the BK equation with fixed $\alpha_s=0.1$ (dashed)
with the initial condition of set g1118.
Right: the same with replacing $Q_{s0}^2=6 Q_{s0,p}^2$.}
\label{fig:BKsol}
\end{figure}

In Fig.~\ref{fig:updf}, 
we show the profile of uGDs,
$\varphi_{p,y}(\k_\perp)$ and $\phi_{A,y}(\k_\perp)$,
of the proton and the nucleus, respectively, 
at several values of $y=\ln(x_0/x)$,
obtained by solving the rcBK equation 
with set g1118 in Table~\ref{tab:par}.
We also show the result of the BK evolution with the fixed coupling
constant $\alpha_s=0.1$, for comparison. 
This small coupling is necessary to keep 
the evolution speed compatible with the empirical value
in the BK equation. 
We see that with increasing rapidity $y$ 
the peak position ({\it i.e.,} the saturation scale)
drifts slightly faster in the rcBK evolution than 
that in the BK evolution with $\alpha_s=0.1$,
and that the former yields a steeper $k_\perp$-slope than the latter.
Note that the uGD should behave as $1/k_\perp^2$
at large $k_\perp$ as is known in the LO double log approximation for 
BFKL or DGLAP equation, while the BFKL (equivalently BK in the linear regime)
evolution gives harder $k_\perp$ spectrum.
A dip structure around $k_\perp=3$ GeV at $x=x_0$ is caused 
by the parameter $\gamma>1$, but is soon
smeared out in the evolution.
In the nucleus case, we assumed the initial saturation scale
$Q_{s0,A}^2 = 6 Q_{s0,p}^2$ at $x=x_0=0.01$. 
The uGD is more suppressed in low $k_\perp$ region and 
the peak position locates at larger $k_\perp$
than in the proton case, reflecting the stronger multiple
scatterings in the nuclear target.

\begin{figure}[tbp]
\centering
\resizebox*{!}{5.5cm}{\includegraphics[angle=270]{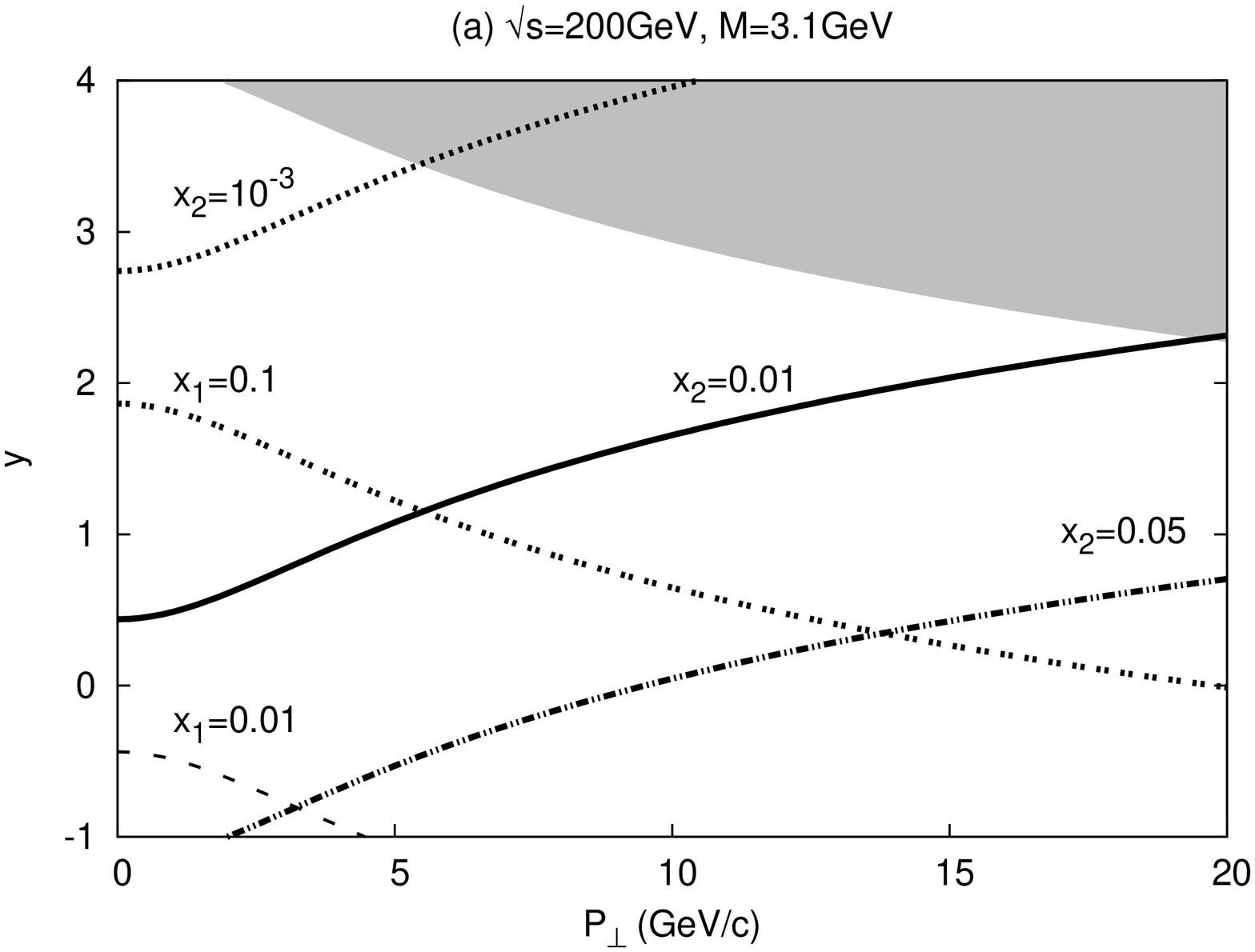}}
\resizebox*{!}{5.5cm}{\includegraphics[angle=270]{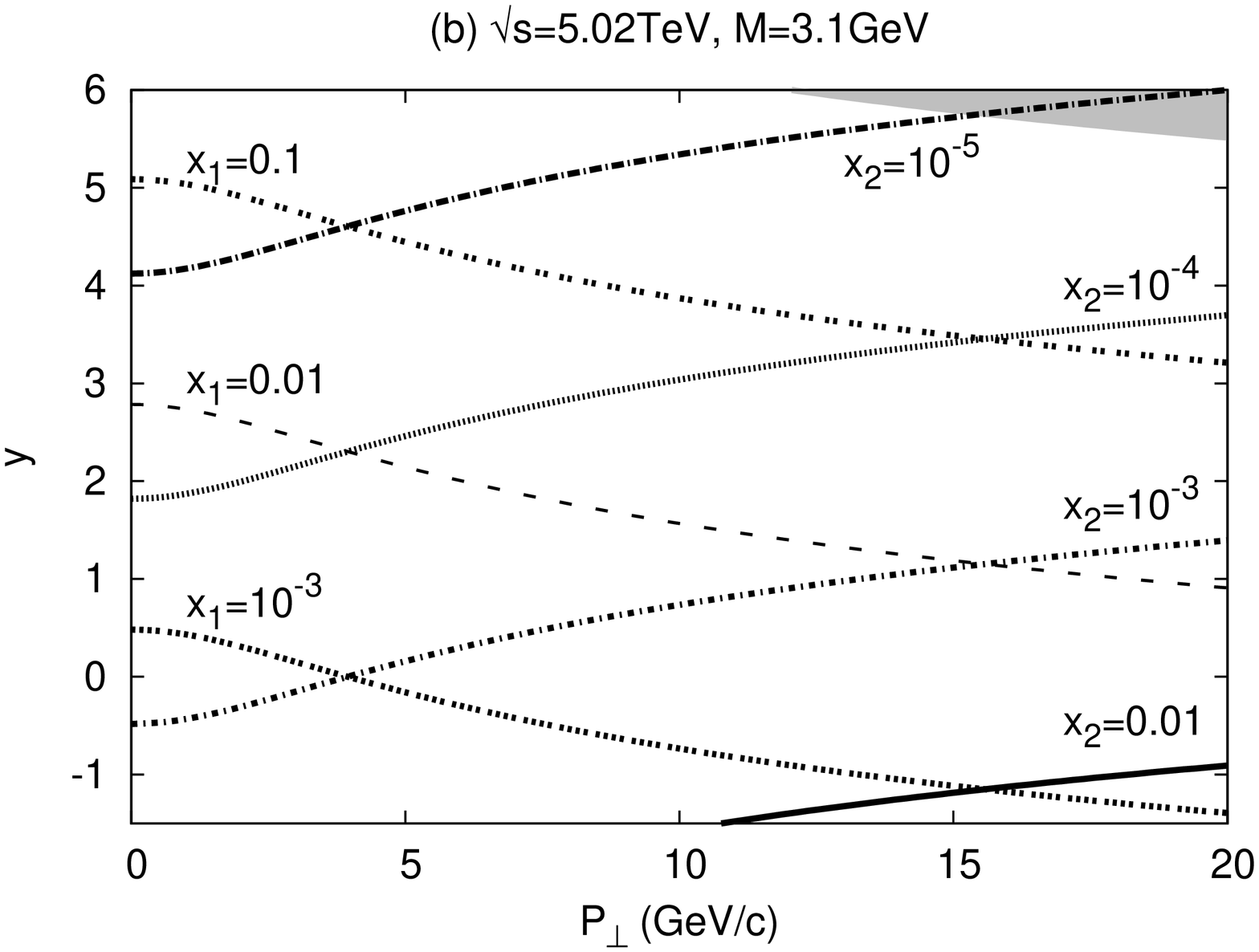}}
\caption{Kinematical coverage of the pair production in
the plane of rapidity $y$ and transverse momentum $P_\perp$
for invariant mass $M=3.1$ GeV  at (a) $\sqrt s$=200 GeV and (b) $\sqrt s$=5.02 TeV.
Shown are the curves of constant $x_{1,2}=(\sqrt{P_\perp^2 +M^2}/\sqrt s)e^{\pm y}$.
The shaded region is kinematically forbidden.}
\label{fig:x2-coverage}
\end{figure}

We consider here the $x_{1,2}$ coverage of the charm pair production
in the plane of the rapidity $y$ and
the transverse momentum $P_\perp$ of the pair
at collision energies $\sqrt{s}$=200 GeV and 5.02 TeV
in Fig.~\ref{fig:x2-coverage}.
Here we fix the pair's invariant mass $M=3.1$ GeV,
and draw the curves determined by
$x_{1,2} = e^{\pm y} (\sqrt{P_\perp^2 +M^2}/\sqrt s)$,
on which either $x_1$ or $x_2$ is constant. 
The kinematically disallowed region where $x_{1,2}>1$ is
indicated by the shaded area.
We see that, at the RHIC energy, 
J/$\psi$ is produced 
from the gluons of moderate $x_{1,2}\sim 0.01-0.05$
at mid-rapidities, 
while at forward rapidities $y\sim 2$
the process gets sensitivity to the gluons at small $x_2<0.01$.
At the LHC energy, on the other hand, J/$\psi$ production
is already sensitive to the small $x_2$ gluon even at mid-rapidity,
and at forward rapidity it probes $x_2$ as low as $\sim 10^{-(4-5)}$.

In the next section, we study the quarkonium production in CEM
applied for the heavy-quark production cross-section 
(\ref{eq:cross-section-LN}) or
(\ref{eq:cross-section-LN-coll})
with the $x$-evolved uGD $\phi_{A,Y}^{q\bar q,g}(\k)$.
CEM has been successful in describing the J/$\psi$ production
in high-energy proton-proton (pp) collisions.

\section{Quarkonium production}
We estimate the J/$\psi$ production
from the quark-pair production cross section 
within CEM\cite{Brambilla:2004wf}:
\begin{align} 
\frac{d\sigma_{\text J/\psi}}{d^2\P_{\perp}dy}
=
F_{\text J/\psi} \; \int_{4m_c^2}^{4M_D^2} dM^2
\frac{d\sigma_{c\bar c}}
{d^2\P_{\perp}dM^2dy}
\, ,
\label{eq:CEM}
\end{align}
where $m_c$ ($M_D$) is the charm quark ($D$-meson) mass.
A phenomenological constant $F_{\text J/\psi}$ represents the 
nonperturbative transition rate 
for the charm pairs, produced in the invariant mass range 
from $2m_c$ to the threshold $2M_D$, to bound into a quarkonium.
Its empirical value is around $F_{\text J/\psi}=0.01$--0.05\cite{Arleoa1}.
Use of CEM for pA collisions 
assumes that the bound state formation occurs 
outside the target nucleus. 
We fix the threshold with
$M_D$ ($M_B$) = 1.864 (5.280) GeV for J/$\psi$ ($\Upsilon$).

A remark is here in order. 
In the pair-production cross section 
(\ref{eq:cross-section-LN})
the inelastic cross section estimated as $\pi R_\text A^2$ 
in the denominator effectively
cancels out with the same factor in $\phi_{A,y}^{q\bar q,g}$,
and that the cross section is now proportional to the
effective transverse area $\pi R_p^2$ of the proton
appearing in $\varphi_{p,y}$.
In the following calculations, we choose 
the proton size $R_\text p=0.9$ fm and
the J/$\psi$ formation fraction $F_{\text J/\psi}=0.02$
as representative values. One should keep in mind that
the absolute normalization of the cross section
depends on these parameters.
We also cancels $\alpha_s$ in front of the cross section
by $\alpha_s$ appearing in the denominator in $\phi_{A,y}$ and
$\varphi_{p,y}$. 
In the case of collinear approximation on the proton side,
we set $\alpha_s=0.2$ in this paper.

\subsection{Transverse momentum spectrum of J/$\psi$}

\subsubsection*{RHIC}

\begin{figure}[tbp]
\begin{center}
\resizebox*{!}{5.5cm}{\includegraphics[angle=270]{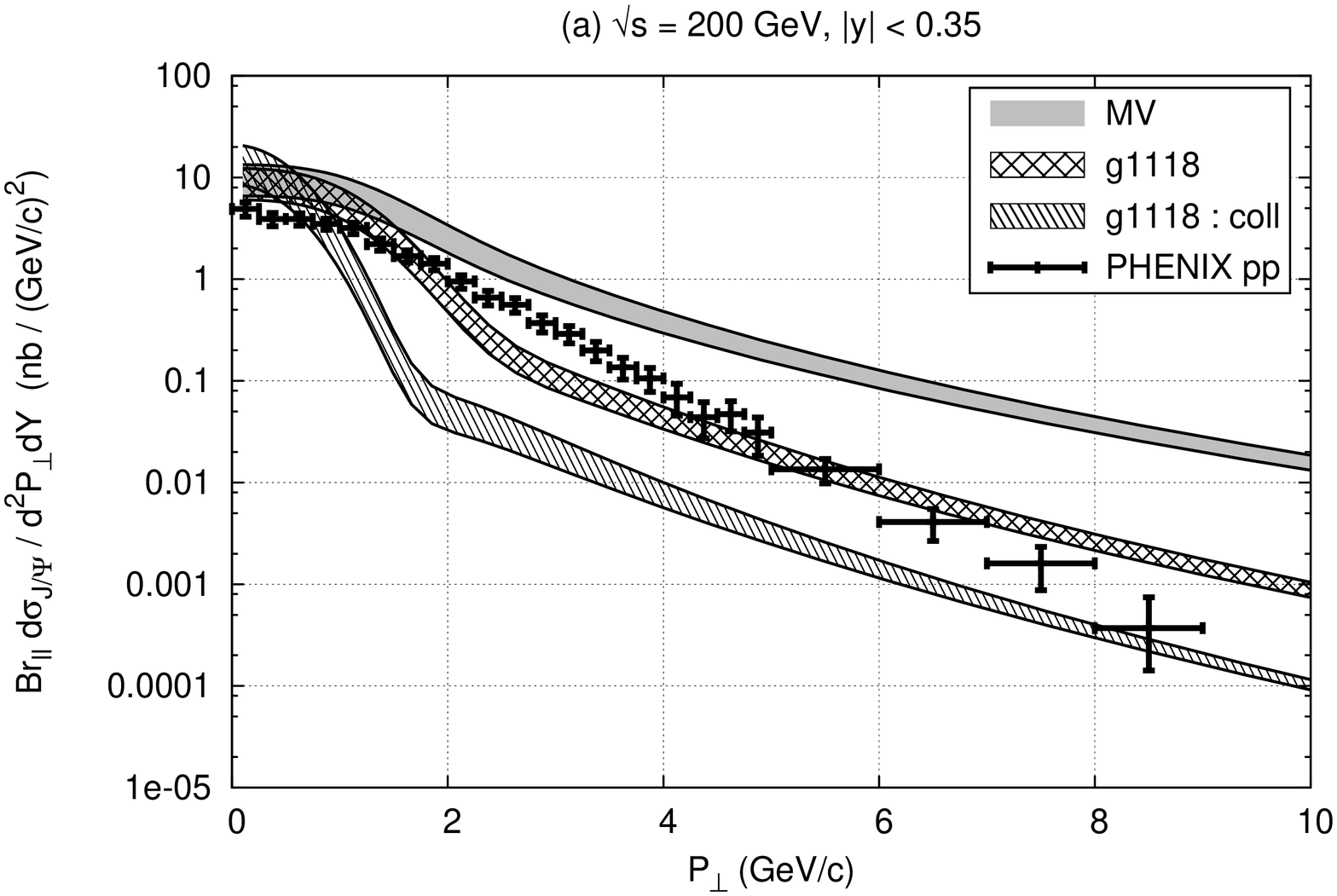}}
\resizebox*{!}{5.5cm}{\includegraphics[angle=270]{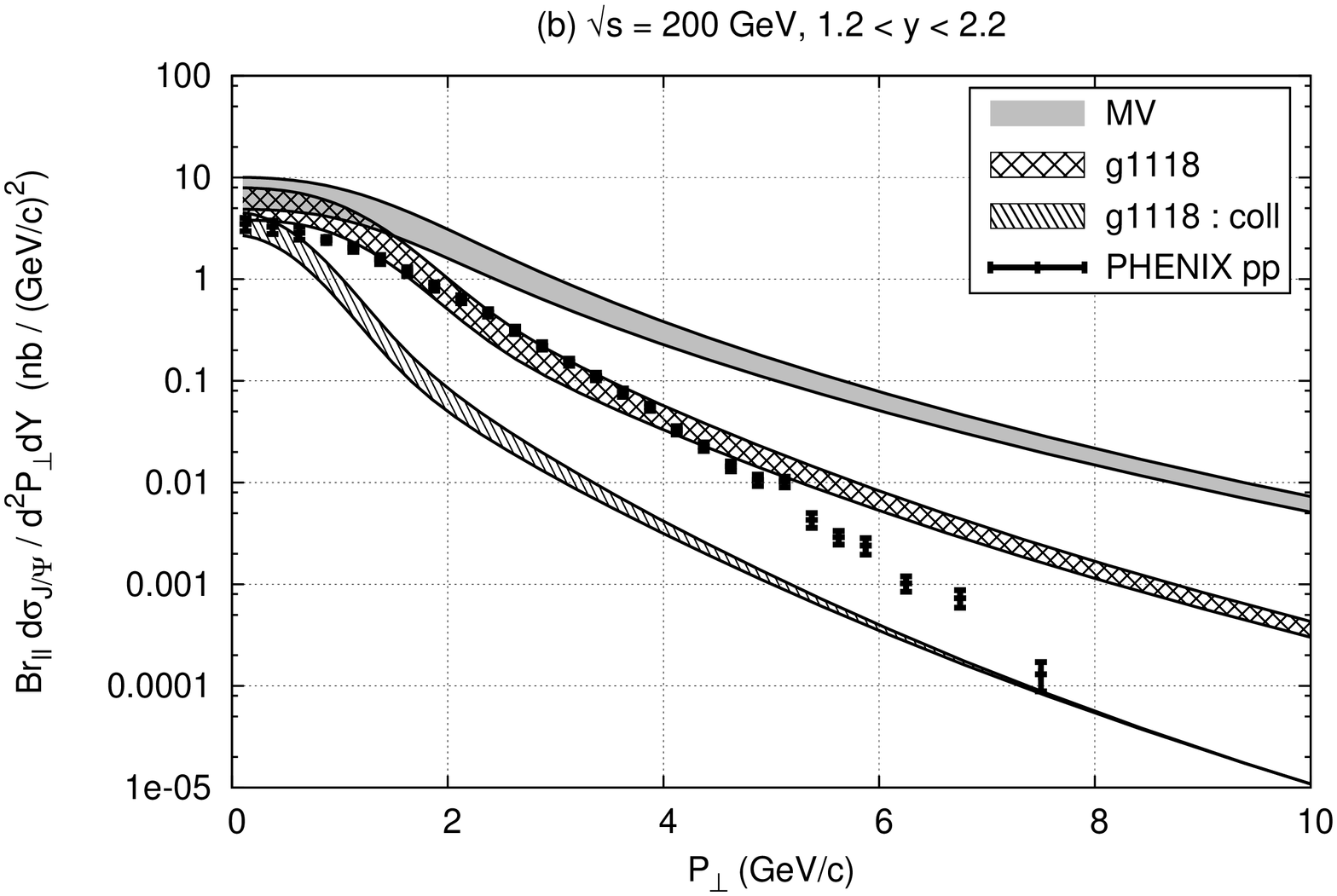}}
\end{center}
\caption{
Transverse momentum spectrum of J/$\psi$ 
in di-lepton channel in pp collisions
at $\sqrt{s}=200$ GeV for rapidity ranges 
(a) $|y|<0.35$ and (b) $1.2<y<2.2$.
CEM model results using the pair production~(\ref{eq:cross-section-LN})
with sets MV and g1118 are shown in gray and doubly-hatched bands,
respectively,
and the result using collinear 
approximation~(\ref{eq:cross-section-LN-coll}) with set g1118 is in
hatched band.
The upper (lower) curve of the band corresponds to 
the result with $m_c=1.2$ (1.5) GeV,
and the scale of pdf is chosen at $2M_\perp$ ($M_\perp/2$) in the
collinear approximation.
Data from~\cite{Adare2}.
}
\label{fig:Jpsi_pt_rhic_pp}
\end{figure}

We first show in Fig.~\ref{fig:Jpsi_pt_rhic_pp}
the transverse momentum spectrum of the
produced J/$\psi$ in pp collisions 
\footnote{Strictly speaking, 
treating pp (at mid-rapidity) as a dilute-dense system is not
legitimate but we need the pp cross sections
for studying the so-called nuclear modification factor 
$R_\text{pA}$ (see text).}
at $\sqrt{s}=200$ GeV, using  
the uGD set g1118 given in Table~\ref{tab:par}.
The upper (lower) curve of each band indicates the
result with charm quark mass $m_c=1.2$  (1.5) GeV.
In the collinear approximation on the larger-$x_1$ side,
we adopt CTEQ6LO parametrization\cite{CTEQ6}, 
and the band in Fig.~\ref{fig:Jpsi_pt_rhic_pp}
includes the change of
the factorization scale from $2M_\perp$ to $M_\perp/2$
with $M_\perp=(M^2+P_{\perp}^2)^{1/2}$,
where $M$ is the pair's invariant mass.

As mentioned above, the quarkonium production
at mid-rapidity $|y|<0.35$ is largely determined by the gluon
distributions at moderate $x_{1,2} \gtrsim 0.01$. 
Then, we notice a difficulty with set g1118:
the peculiar dip structure of g1118 seen in Fig.~\ref{fig:BKsol}
remains in the J/$\psi$ spectrum
as a similar dip around $P_\perp \sim 2$ GeV, 
which must be an artifact of this initial condition. 
In contrast, 
we don't see such a structure with the MV initial condition.
At forward-rapidity $1.2<y<2.2$, 
the dip is smeared to be less noticeable
by the imbalance between $x_1$ and $x_2$ and
by the $x_2$ evolution of the uGD.
As a whole, 
the $P_\perp$ spectrum obtained with set g1118 is closer
to the observed data~\cite{Adare2} than with set MV. 
In this pp case, the collinear approximation
on the large-$x_1$ side does not improve
the description of the data. 
The $k_\perp$ kick from only the one of the protons cannot 
give enough $P_\perp$ for the pair.

\begin{figure}[tbp]
\begin{center}
\resizebox*{!}{5.5cm}{\includegraphics[angle=270]{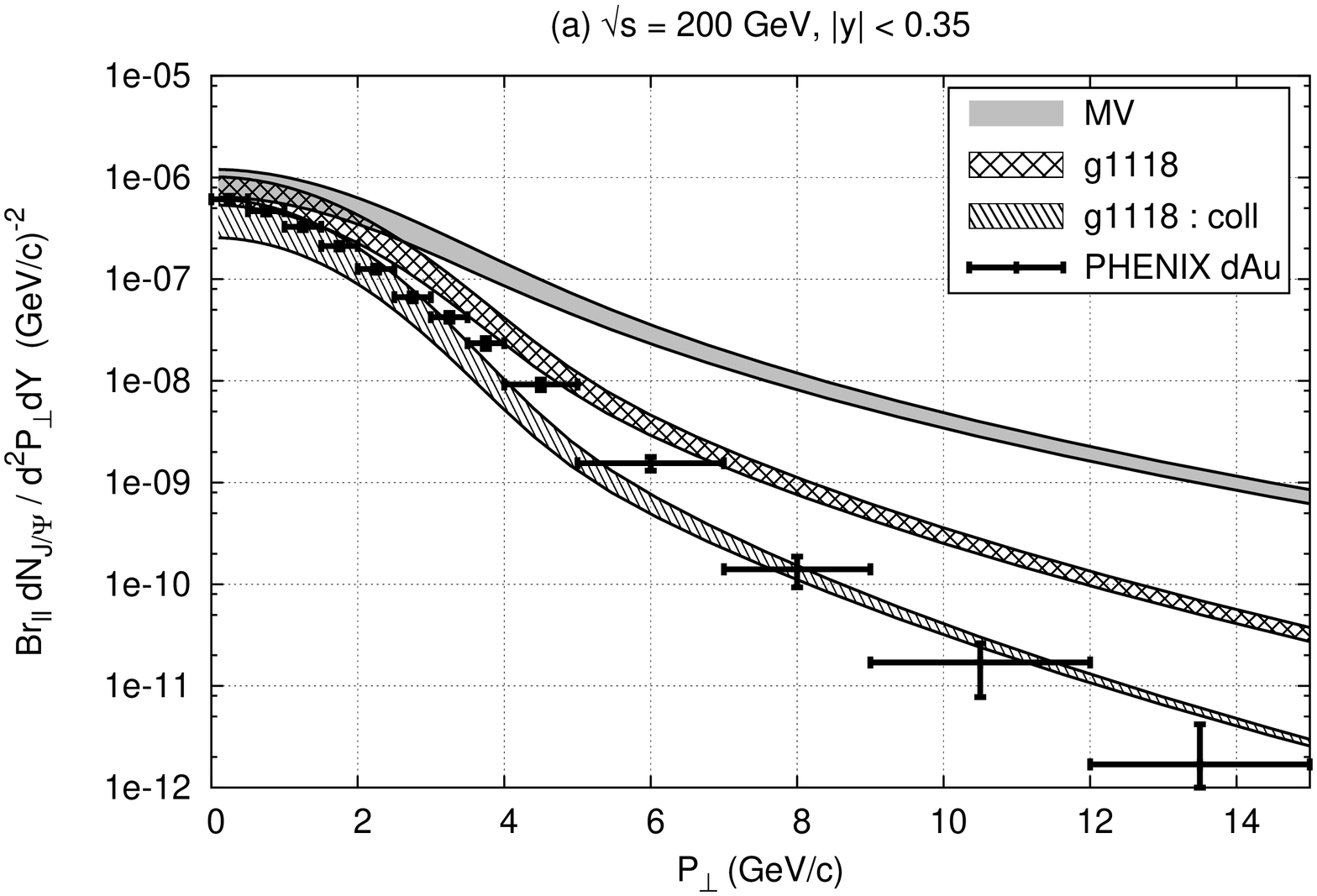}}
\resizebox*{!}{5.5cm}{\includegraphics[angle=270]{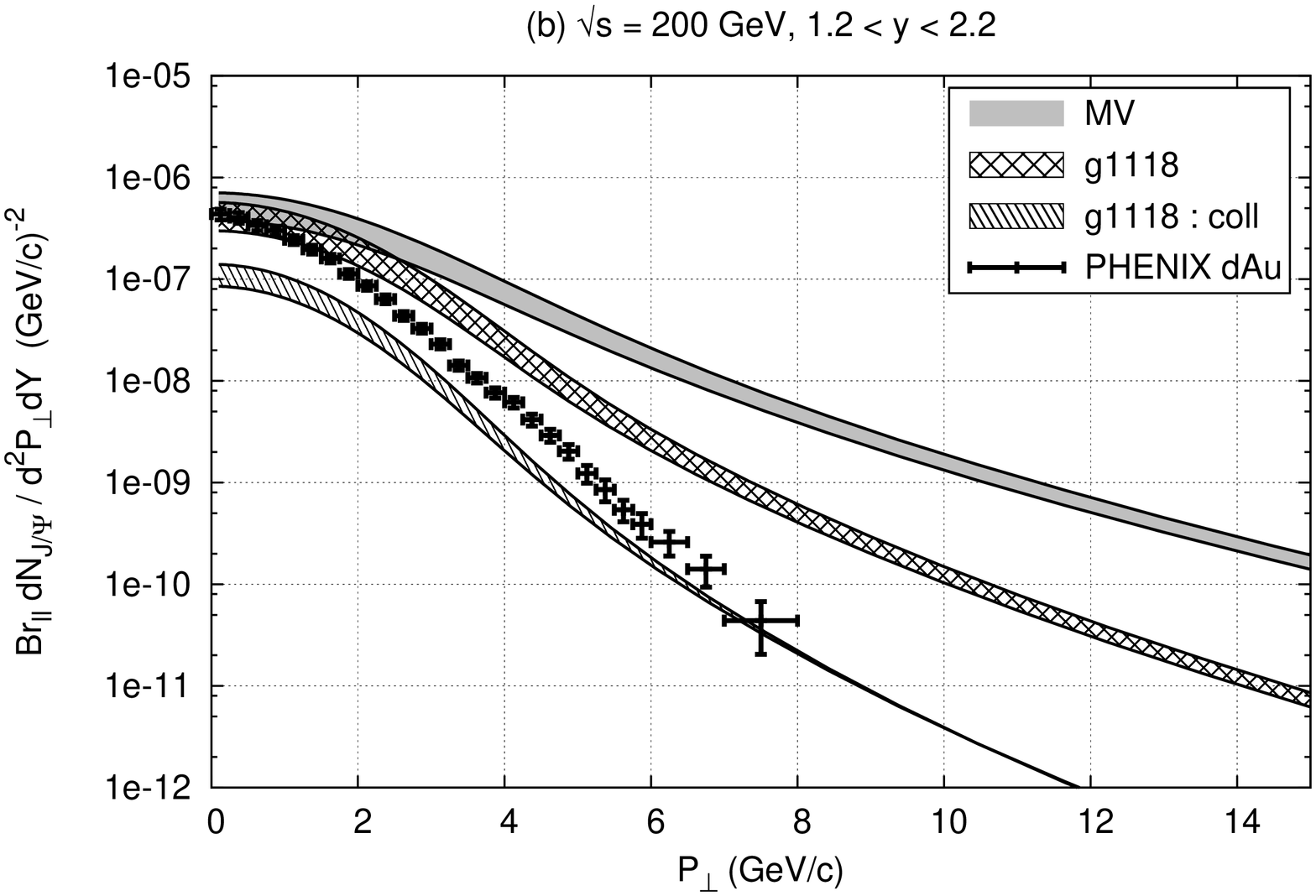}}
\end{center}
\caption{
Transverse momentum spectrum of J/$\psi$ 
in di-lepton channel in pA collisions
at $\sqrt{s}=200$ GeV
for rapidity ranges 
(a) $|y|<0.35$ and (b) $1.2<y<2.2$.
Notations are the same as in Fig.~\ref{fig:Jpsi_pt_rhic_pp}.
Data in d-Au collisions\cite{Adare3} are overlaid for comparison.}
\label{fig:Jpsi_pt}
\end{figure}

In Fig.~\ref{fig:Jpsi_pt} shown is the 
transverse momentum spectrum of the J/$\psi$ in pA collisions
in our model. 
We set the initial saturation scale of the uGD 
for the heavy nucleus as $Q_{s0,A}^2=6 Q_{s0,p}^2$ at $x=x_0$. 
The upper (lower) curve of each band indicates the result
with $m_c=1.2$ (1.5) GeV.
We overlay d-Au data observed by PHENIX at $\sqrt{s}=200$ GeV\cite{Adare3},
presuming here that the difference between pA and dA
results only in normalization difference of order O(1) 
\footnote{Recall that our model already has an uncertainty of 
O(1) in the normalization of the uGD.}.
We find that $P_\perp$-dependence of J/$\psi$ production
is better described with set g1118~\footnote{%
Possible dip structure from the proton uGD is smeared out here 
in Fig.~\ref{fig:Jpsi_pt} by the multiple scattering effects
in the nuclear uGD.} than that with set MV.
Indeed, here the collinear approximation on the proton side
apparently gives a better description of the data both at mid- and 
forward-rapidity regions. 
However, at forward rapidities,
where we approach the small-$x_2$ region and 
the kinematical boundary for $x_1$ at the same time 
(see Fig.~\ref{fig:x2-coverage}),
we expect a nontrivial interplay between 
large $x_1$ and small $x_2$.
Besides the saturation dynamics of $x_2$ gluons, one may need to 
consider other physics such as energy loss of large-$x_1$ gluons 
in the heavy target\cite{Arleo:2012rs} in order to understand the
$P_\perp$ spectrum of J/$\psi$ in the very forward region.
These effects are not included in our present treatment.

We notice in Fig.~\ref{fig:Jpsi_pt} (b)
that the J/$\psi$ production is more suppressed nearly by one order of
magnitude in the collinear approximation than
those in the full calculation.
This is caused by a difference in the large $x_1$ behavior of
gluon distributions on the proton side. 
As $x \to 1$, the CTEQ gluon distribution decreases
much more rapidly than our model uGD $\varphi_{p,y}$,
which is assumed as $\propto (1-x)^4$. 
Furthermore, 
in the collinear approximation, the pair's $P_\perp$
is entirely provided from the nucleus side, $P_\perp = k_2$, 
and uGD $\phi_{A,y}$ for the heavy target is more suppressed
at low $k_2$ by multiple scatterings.

\begin{figure}[tbp]
\begin{center}
\resizebox*{!}{5.5cm}{\includegraphics[angle=270]{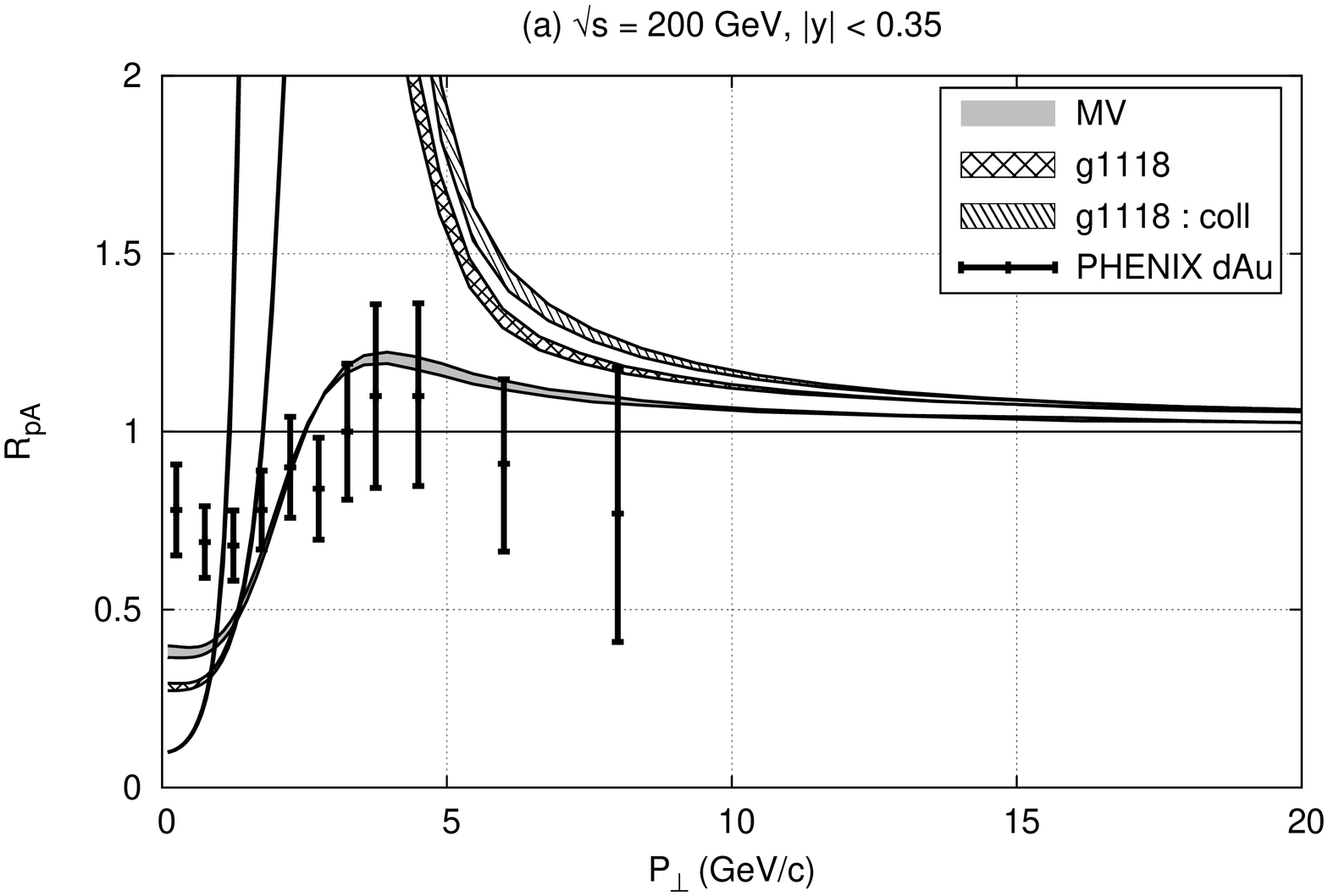}}
\resizebox*{!}{5.5cm}{\includegraphics[angle=270]{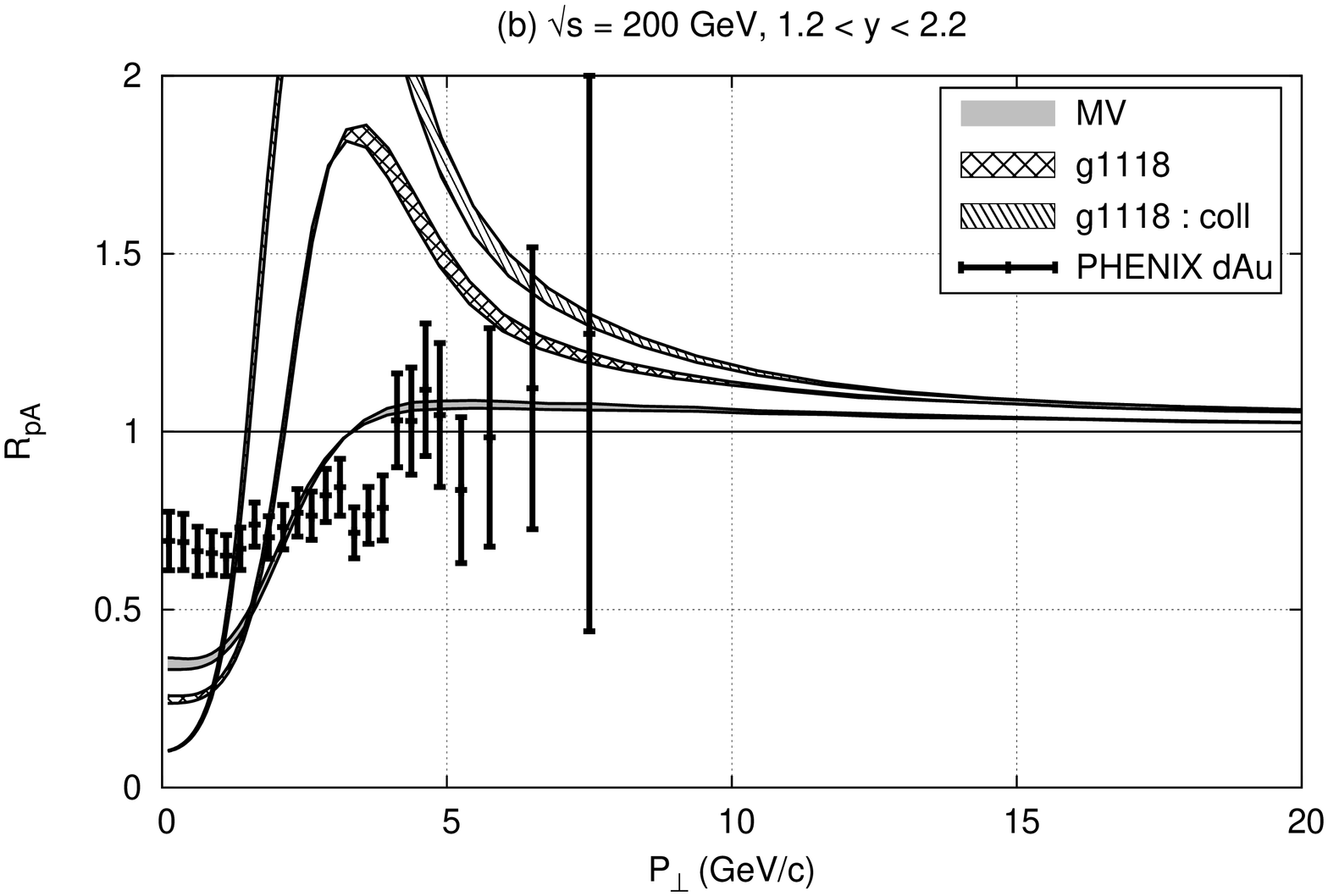}}
\end{center}
\caption{The ratio of J/$\psi$ productions in pA and pp collisions $R_\text{pA}(P_\perp)$
at $\sqrt{s}=200$ GeV for (a) $|y|<0.35$ and (b) $1.2<y<2.2$.
The results with uGD sets MV and g1118 are shown in gray and doubly-hatched
bands, respectively, and the result in collinear approximation
with set g1118 is shown in a hatched band.
Notations are the same as in Fig.~\ref{fig:Jpsi_pt_rhic_pp}.
Data of $R_{dAu}$ taken from \cite{Adare3}.}
\label{fig:RpA_Jpsi_rhic}
\end{figure}

Now let us take a ratio of the cross section of J/$\psi$ 
in pA collisions to that in pp collisions, 
which is called nuclear modification factor $R_\text{pA}$.
We expect that model uncertainties cancel out to some extent in the
ratio.
We define $R_\text{pA}$  for J/$\psi$ in our model as 
\begin{align}
R_\text{pA}= \frac{\left . d\sigma_{{\rm J}/\psi}/d^2 P_\perp dy \right |_\text{pA}}
  {N_\text{coll}\, \left . d\sigma_{{\rm J}/\psi}/d^2 P_\perp dy \right |_\text{pp}}
\; .
\label{eq:def-RpA}
\end{align}
Here we set the number of nucleon-nucleon collisions in pA 
to $N_\text{coll}=A^{\gamma / 3}$
because the uGD $\phi_{A,y_0}(\k_\perp)$ scales 
as $(Q_{s0}^2)^{\gamma}\propto A^{\gamma/3}$ at large $k_\perp$.

In Fig.~\ref{fig:RpA_Jpsi_rhic} we compare the model results for 
$R_\text{pA}$ at $\sqrt s=200$ GeV with the data of $R_\text{dAu}$.
Note that the projectile is different between the model calculation and 
the data.
The notations are the same as in Fig.~\ref{fig:Jpsi_pt_rhic_pp}.
We stress here that $R_{\rm pA}$ is indeed little dependent on the choice
of the quark mass and factorization scale.
Unfortunately, however, one immediately recognizes an unphysically strong
Cronin peak in the model calculations with set g1118
both at mid- and forward rapidities, which is
obviously caused by the dip seen in the pp collisions
(Fig.~\ref{fig:Jpsi_pt_rhic_pp}).
In contrast, the $R_{\rm pA}$ result with set MV looks more reasonable;
we see a moderate Cronin peak at mid-rapidity
due to the multiple scatterings, 
while it almost disappears at forward rapidity $y \sim 2$ 
by the $x_2$ evolution.
In low-$P_\perp$ region, we also notice a too strong suppression 
than the experimental data. 
This would imply the importance of the fragmentation process in
the formation of J/$\psi$, which is missing 
in a simple CEM treatment.

To summarize the results at RHIC energy,
the J/$\psi$ production spectrum is sensitive
to the moderate value of $x_{1,2}$, where the initial 
condition for the $x$-evolution is set.
We have a difficulty to describe the pp data
and therefore the ratio $R_\text{pA}$ 
with the uGD g1118 constrained at $x<0.01$. 
In contrast the set MV gives more reasonable behavior
for $R_\text{pA}$. 
In pA collisions the $P_\perp$ spectrum  is
better described with set g1118 at mid- and forward-rapidities.
In forward rapidity, the observed $P_\perp$ slope is still
steeper than the model, hinting other effects such as
a possible energy loss of the large-$x_1$ gluon from the proton.
Actually $R_{\rm dA}$ of J/$\psi$ at RHIC energy 
has been studied in several approaches (e.g.)
with introducing nuclear parton distribution and 
nuclear absorption effects
to a J/$\psi$ production model for pp\cite{Ferreiro:2008wc},
or with taking account of 
the multiple scatterings and energy loss of the projectile
gluons\cite{Arleo:2012rs}.

\subsubsection*{LHC}

\begin{figure}[tbp]
\begin{center}
\resizebox*{!}{5.5cm}{\includegraphics[angle=270]{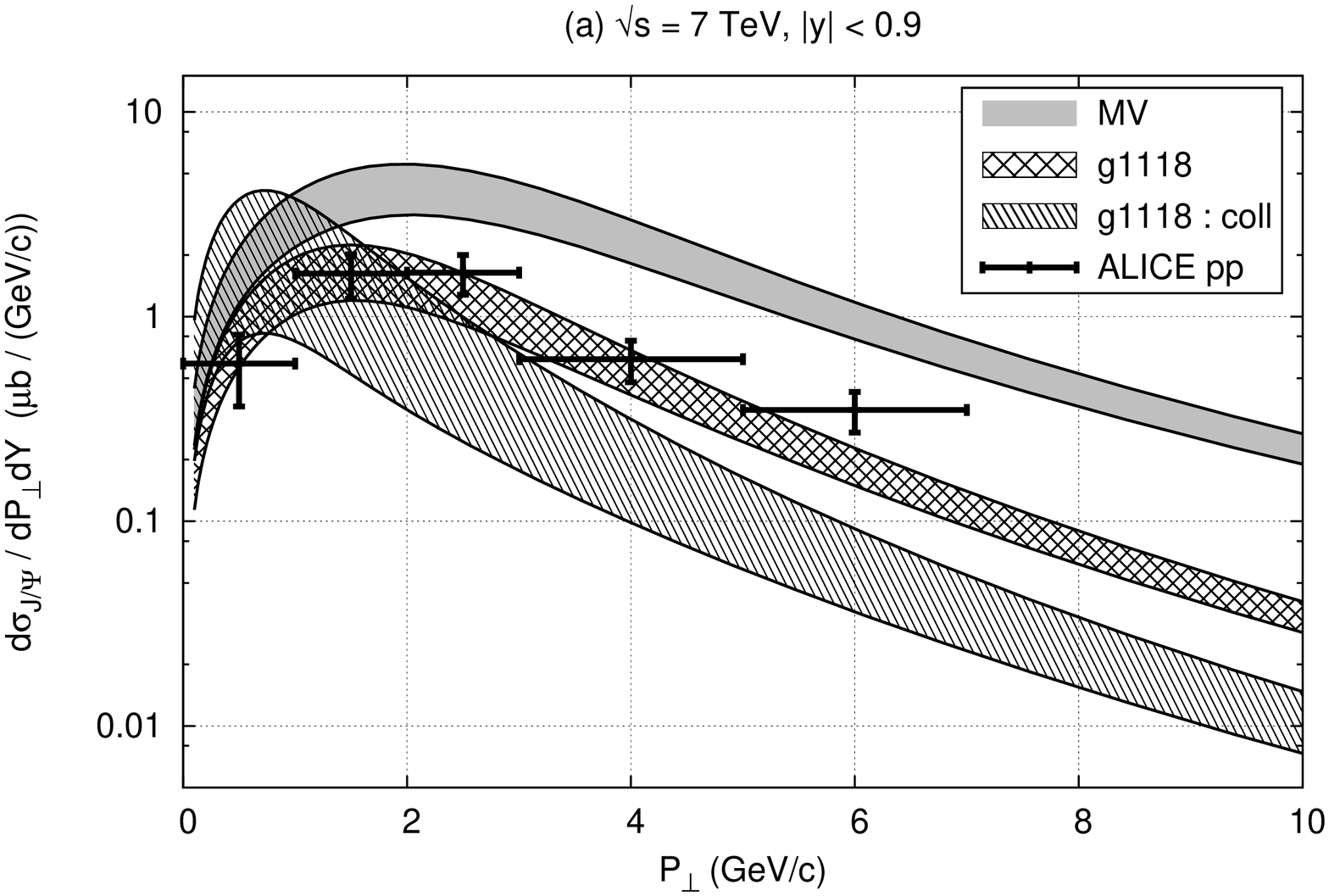}}
\resizebox*{!}{5.5cm}{\includegraphics[angle=270]{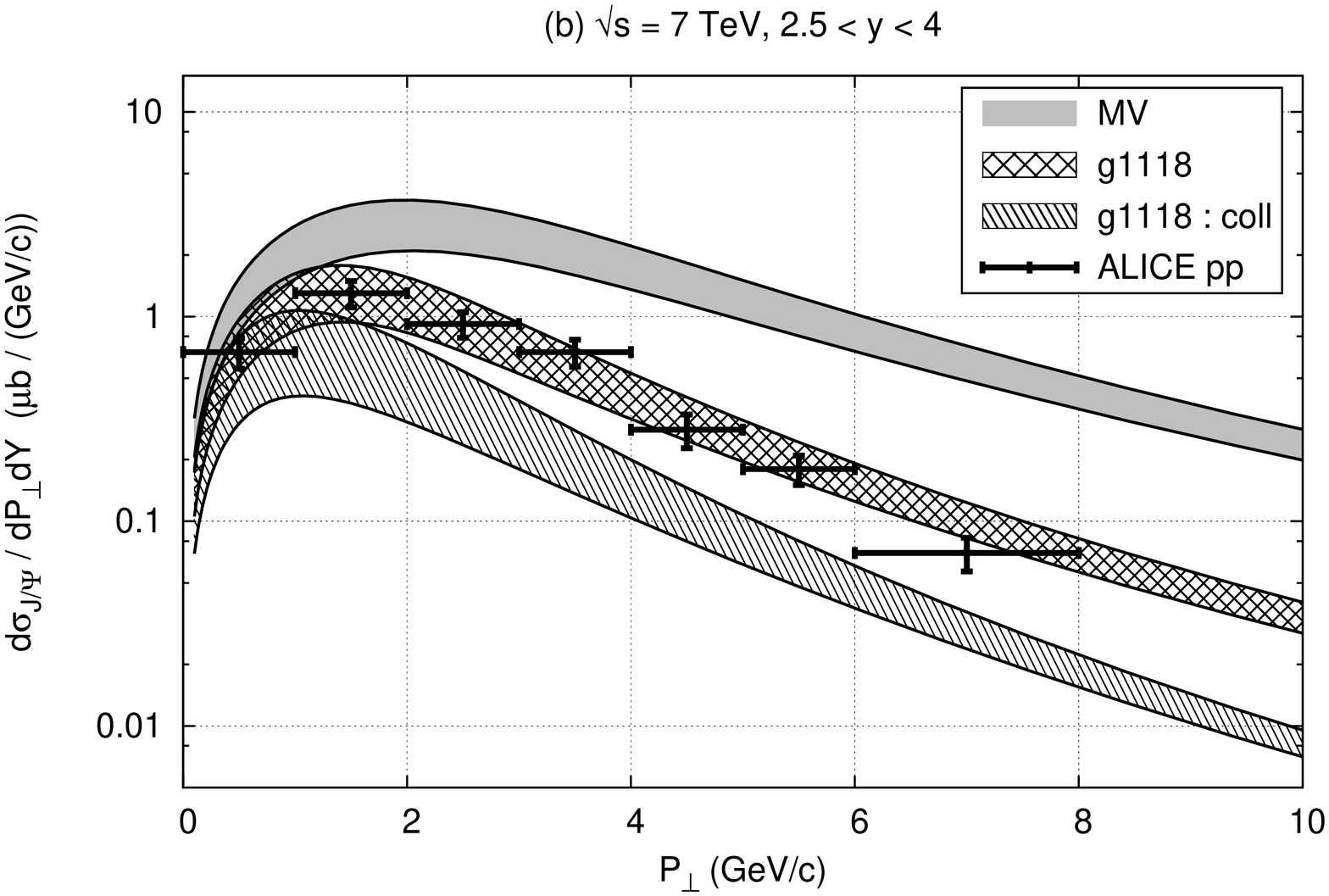}}
\end{center}
\caption{
Differential J/$\psi$ yield in pp collisions
at 
$\sqrt{s}=7$ TeV for (a) $|y|<0.9$ and (b) $2.5<y<4$. 
Notations are the same as in Fig.~\ref{fig:Jpsi_pt_rhic_pp}.
Data from \cite{Aamodt1}.
}
\label{fig:Jpsi-pp-LHC}
\end{figure}

\begin{figure}[tbp]
\begin{center}
\resizebox*{!}{5.5cm}{\includegraphics[angle=270]{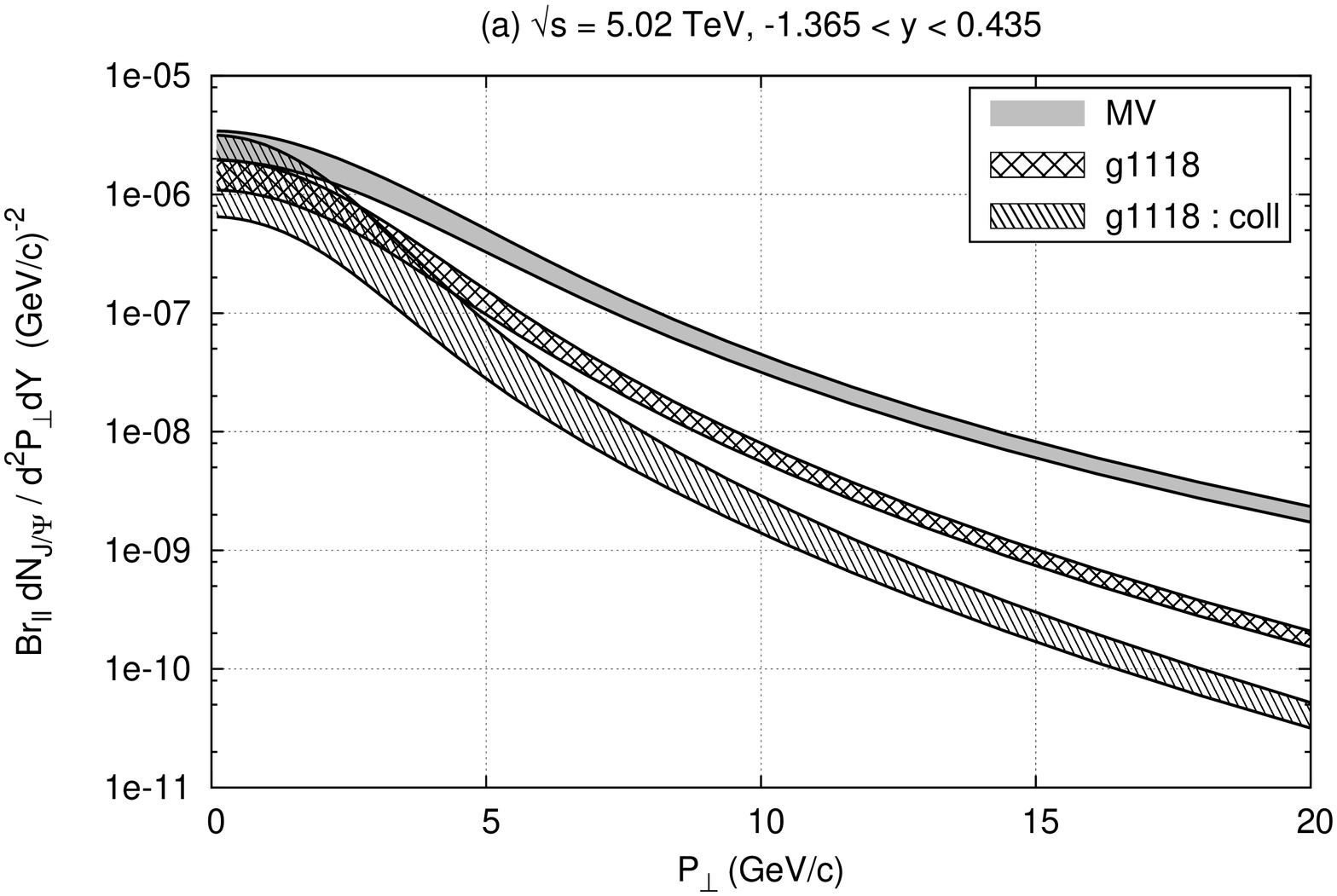}}
\resizebox*{!}{5.5cm}{\includegraphics[angle=270]{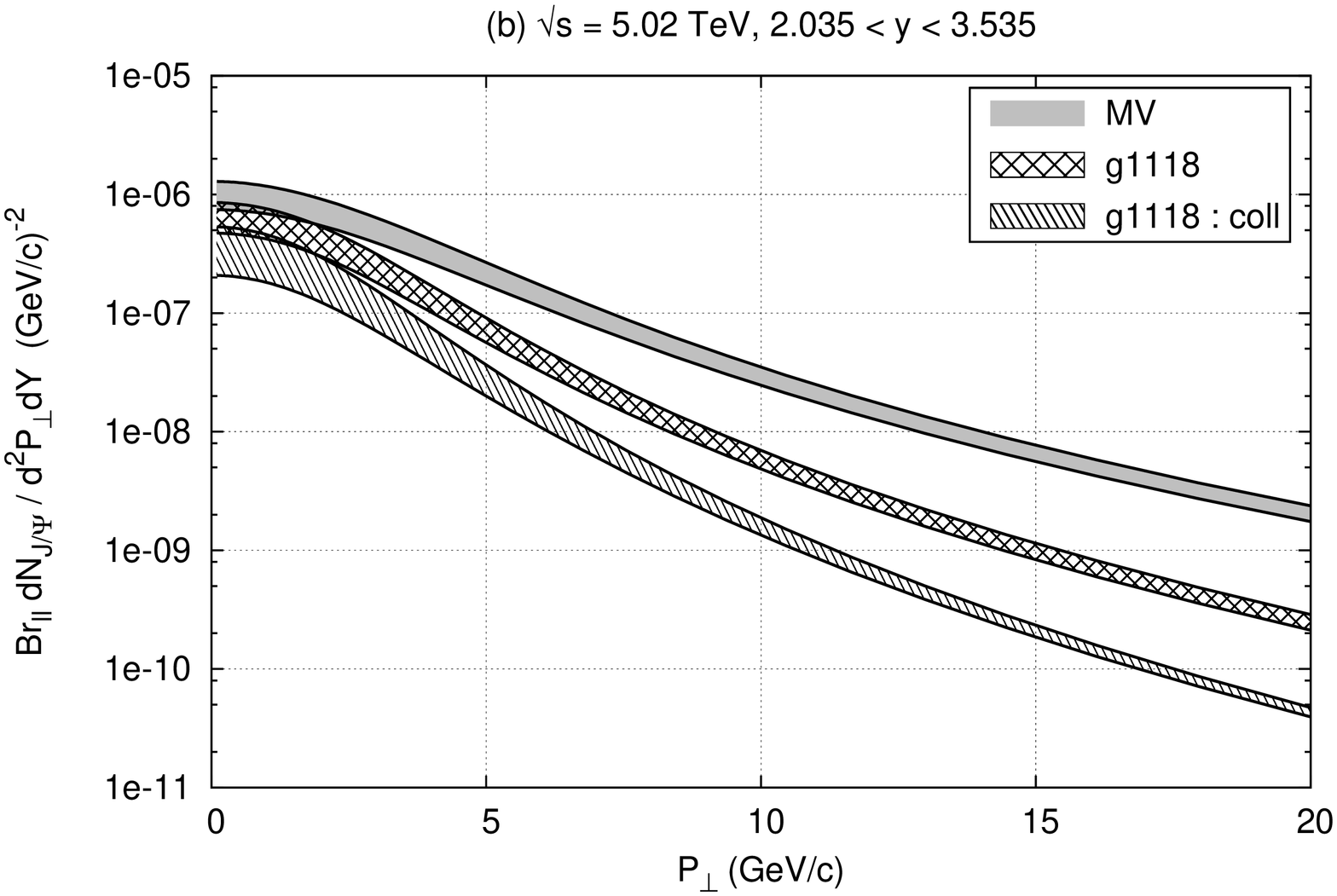}}
\end{center}
\caption{Transverse momentum spectrum of J/$\psi$ 
in di-lepton channel in pA collisions at $\sqrt{s}=5.02$ TeV 
for (a) $-1.4<|y|<0.4$ and (b) $2<y<3.5$.
Notations are the same as in Fig.~\ref{fig:Jpsi_pt_rhic_pp}.}
\label{fig:Jpsi_pt_lhc}
\end{figure}

Now we compute the J/$\psi$ production at the LHC energy,
where we expect that a wider $x_2$-evolution of uGD on the nucleus
side will manifest in the quarkonium spectrum.
In fact, both $x_{1,2}$ are small ($\sim 10^{-3}<x_0$)
already in mid-rapidity production of the charm pair 
as seen in Fig.~\ref{fig:x2-coverage},
and as moving to larger rapidities we can probe
smaller values of $x_2$ on the nucleus side down to $x_2 \sim 10^{-5}$.

We show in Fig.~\ref{fig:Jpsi-pp-LHC} the J$/\psi$ cross section
in pp collision at $\sqrt s=7$ TeV, 
obtained in CEM from charm quark spectrum (\ref{eq:cross-section-LN}).
Notations are the same as in the case of the RHIC energy.
In order to assess the uncertainty,
we again vary the charm quark mass from $m_c=1.2$ to 1.5 GeV,
and change the factorization scale from $2M_\perp$ to $M_\perp/2$
in the collinear approximation.
The observed data\cite{Aamodt1} is fairly well reproduced 
with set g1118 in this $P_\perp$ region both at $|y|<0.9$
and $2.5<y<4$, indicating that $y$-dependence is appropriately
captured by $x$ evolution of uGD.
The $P_\perp$ slope in the 
collinear approximation (\ref{eq:cross-section-LN-coll})
with set g1118 seems to be slightly off the data,
while
the full result with set MV gives harder $P_\perp$ spectrum.
The situation is expected to be similar in pp collisions 
at $\sqrt s=5.02$~TeV.

Results in pA collisions at $\sqrt s=5.02$~TeV are plotted
at mid- and forward-rapidities in Fig.~\ref{fig:Jpsi_pt_lhc}. 
The MV initial condition gives a harder spectrum of J/$\psi$
than g1118. But their
$P_\perp$ slopes become almost the same at $P_\perp \gtrsim 10$ GeV,
hinting the same BFKL tail of uGD generated during the evolution.
Compared to the case at the lower energy $\sqrt{s}=200$ GeV,
the collinear approximation (with set g1118) 
results in the spectral shape rather similar to the full result
at this energy $\sqrt{s}=5.02$ TeV, where 
the collinear approximation on the proton side would be more suitable
since the saturation scale of the nucleus is much larger than that of the
proton: $Q_{s,A}^2(x_2) \gg Q_{s,p}^2(x_1)$, especially in the forward region.

\begin{figure}[tbp]
\begin{center}
\resizebox*{!}{5.5cm}{\includegraphics[angle=270]{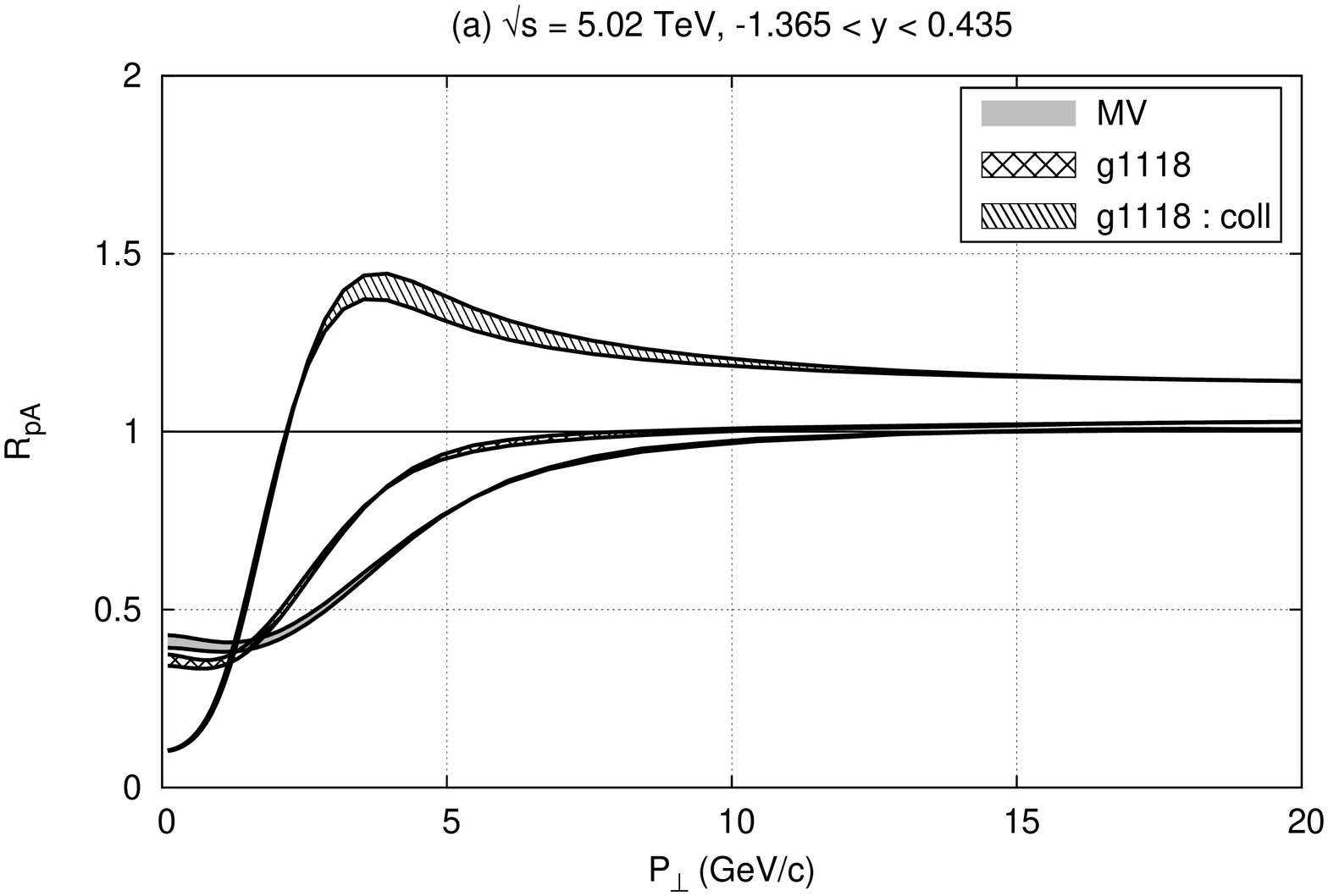}}
\resizebox*{!}{5.5cm}{\includegraphics[angle=270]{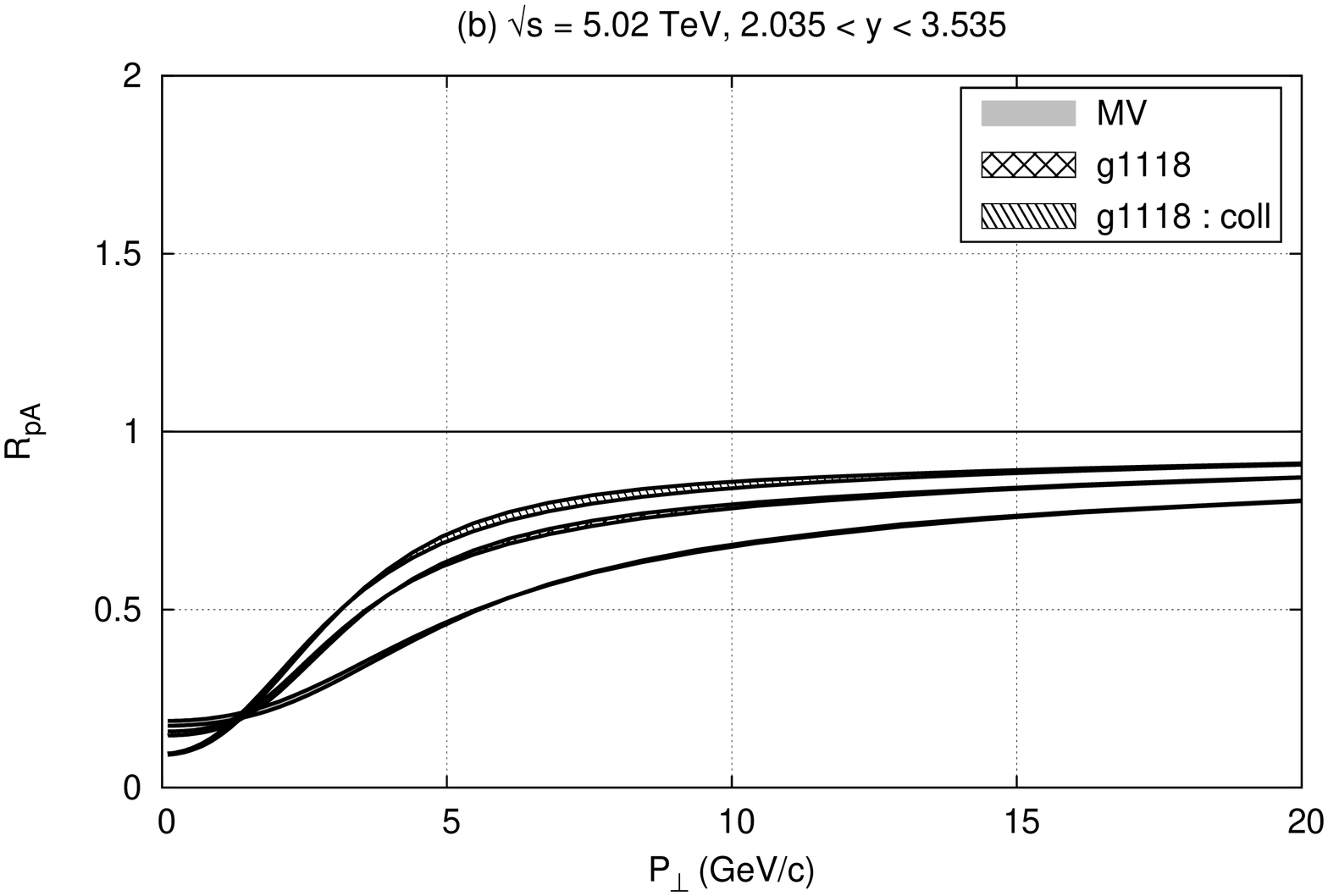}}
\end{center}
\caption{The ratio $R_\text{pA}(P_\perp)$ for J/$\psi$ at $\sqrt{s}=5.02$ TeV 
for (a) $-1.4< y<0.4$ and (b) $2<y<3.5$.
Notations are the same as in Fig.~\ref{fig:Jpsi_pt_rhic_pp}.}
\label{fig:RpA_Jpsi_lhc}
\end{figure}

\begin{figure}[tbp]
\begin{center}
\resizebox*{!}{5.5cm}{\includegraphics[angle=270]{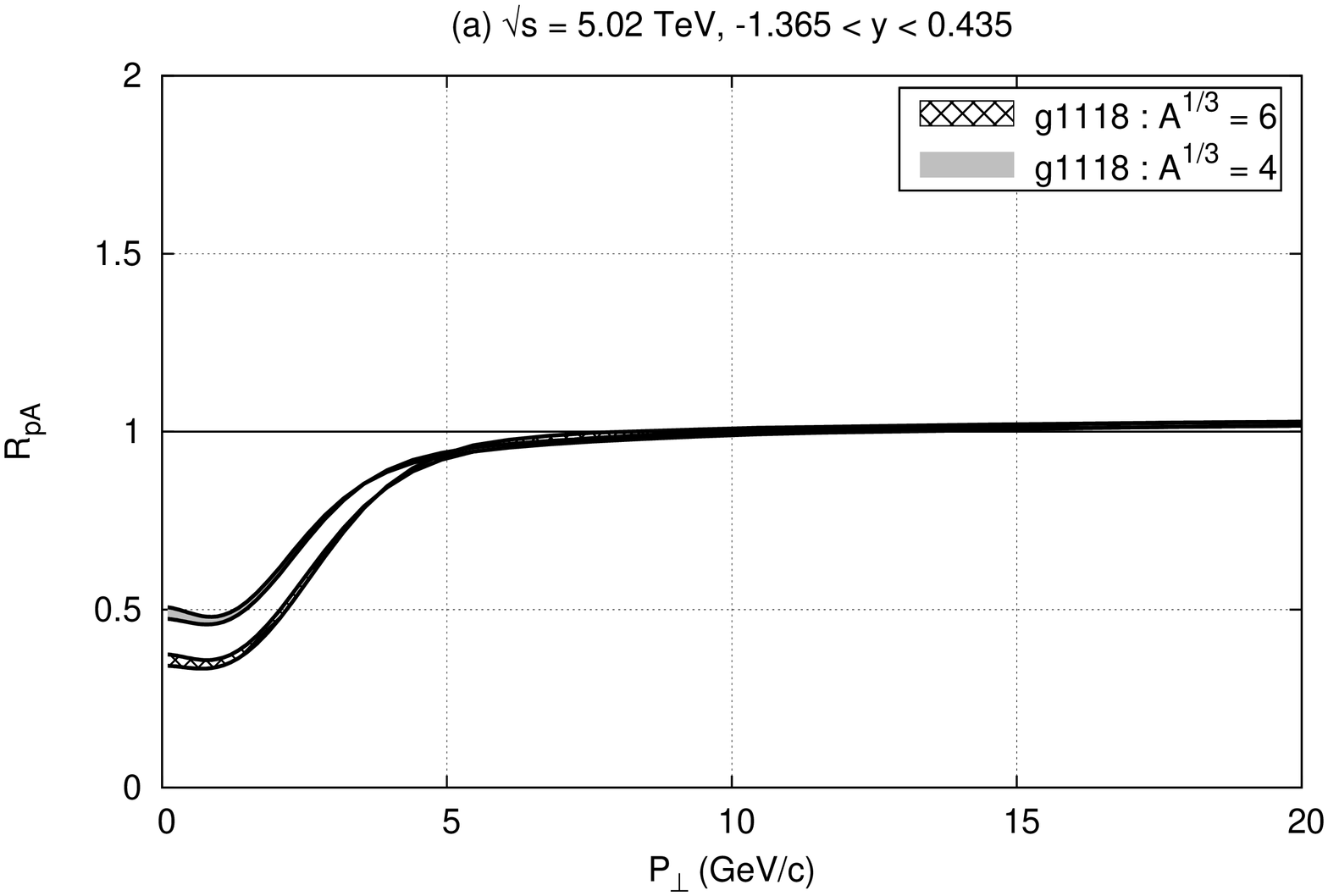}}
\resizebox*{!}{5.5cm}{\includegraphics[angle=270]{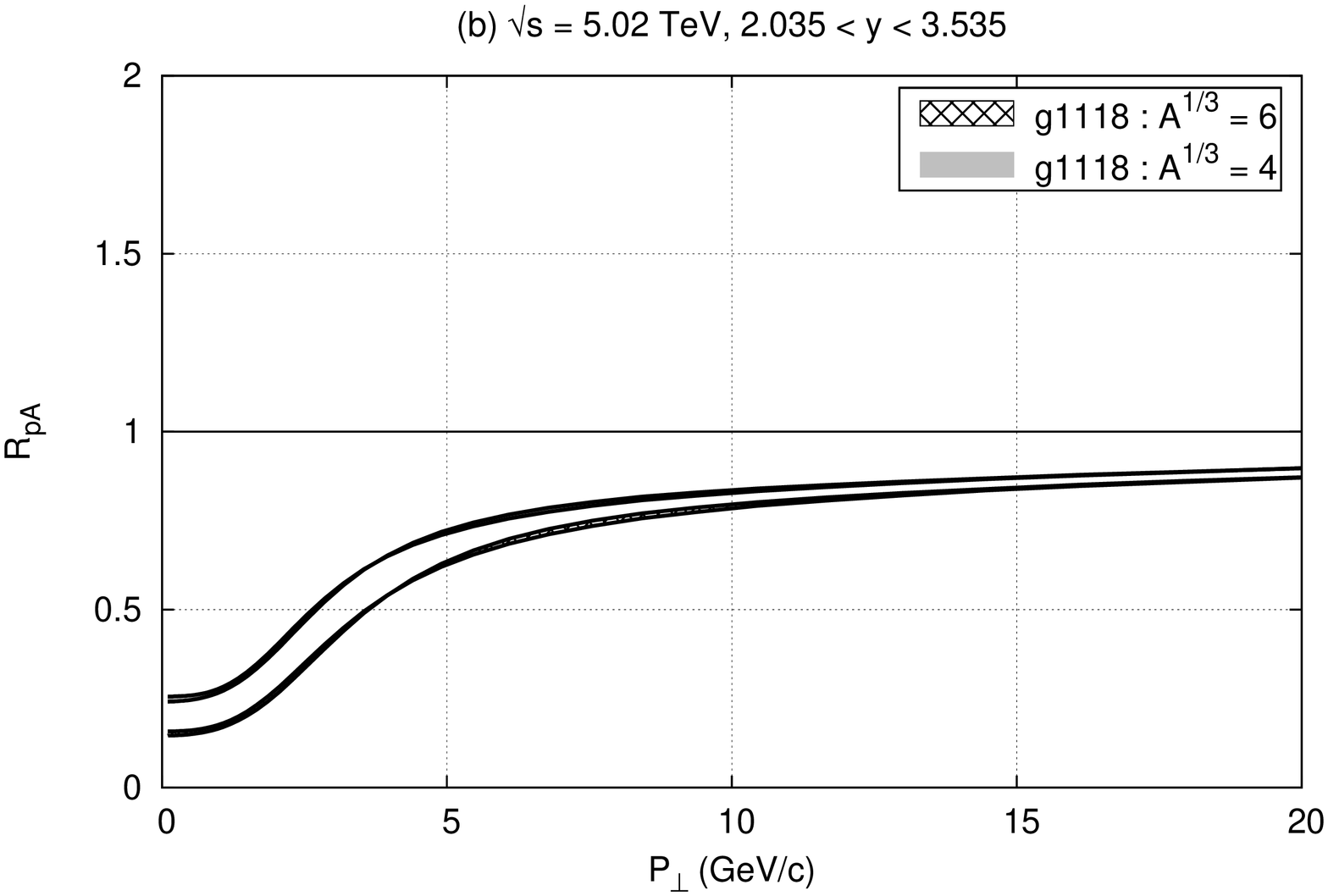}}
\end{center}
\caption{Initial-scale dependence 
of the ratio $R_\text{pA}(P_\perp)$ for J/$\psi$ 
at (a) mid-  and (b) forward-rapidities at $\sqrt{s}=5.02$ TeV. 
$Q_{s0,A}^2$ is set to $4 Q_{s0,p}^2$ (upper) and
$6 Q_{s0,p}^2$ (lower).}
\label{fig:RpA_Jpsi_lhc_46}
\end{figure}

We show in Fig.~\ref{fig:RpA_Jpsi_lhc}
the ratio $R_\text{pA}$ of J/$\psi$
as a function of $P_{\perp}$ at $\sqrt{s}=5.02$ TeV.
We have assumed $N_\text{coll}=A^{\gamma /3}$ as mentioned before.
We find that 
each band almost collapses into a single line, which means that
the ratio $R_\text{pA}$ is insensitive to
the variation of the charm quark mass (and the factorization scale
in the collinear approximation) within the range considered here.

At mid-rapidities (Fig.~\ref{fig:RpA_Jpsi_lhc} (a)), 
we see that the ratio $R_\text{pA}$ of J/$\psi$ production
is suppressed at low $P_\perp$, while it approaches unity 
at higher $P_\perp$ for both sets of g1118 and MV.
In the collinear approximation on the proton side, $R_\text{pA}$
shows a Cronin-like peak around $P_\perp \sim 4$~GeV and remains
larger than unity at larger $P_\perp$, which largely reflects
``$R_\text{pA}$ for $\phi_{A,y}$'' at the gluon level.
At forward rapidities (Fig.~\ref{fig:RpA_Jpsi_lhc} (b)), 
however, this difference due to different uGD sets and approximations
becomes much weaker to yield a systematic suppression as
a function of $P_\perp$ for all three cases.

We examine the initial-scale ($Q_{s0,A}^2$) dependence of
the ratio $R_\text{pA}$ in Fig.~\ref{fig:RpA_Jpsi_lhc_46}, 
by plotting the results with the saturation scale
$Q_{s0,A}^2=4 Q_{s0,p}^2$ (upper) and $6 Q_{s0,p}^2$ (lower) in
Eq.~(\ref{eq:initialcondition}) at $x=x_0$.
It is found that the $Q_{s0,A}^2$ dependence of $R_\text{pA}$ is
relatively weak within this range. At low $P_\perp$ we have strong
suppression, but one should keep in mind that this suppression
may be filled to some extent by the nonperturbative fragmentation
in forming J/$\psi$, as is inferred from the discussion on 
Fig.~\ref{fig:RpA_Jpsi_rhic}.

To summarize the result at LHC energy,
we can probe here a wide $x_2$-evolution of the uGD $\phi_{A,y_2}(\k_2)$
through the J/$\psi$ production, and the ratio $R_\text{pA}$ will be a
good indicator for it.

\subsection{$\Upsilon$ production at the LHC}

\begin{figure}[tbp]
\begin{center}
\resizebox*{!}{5.5cm}{\includegraphics[angle=270]{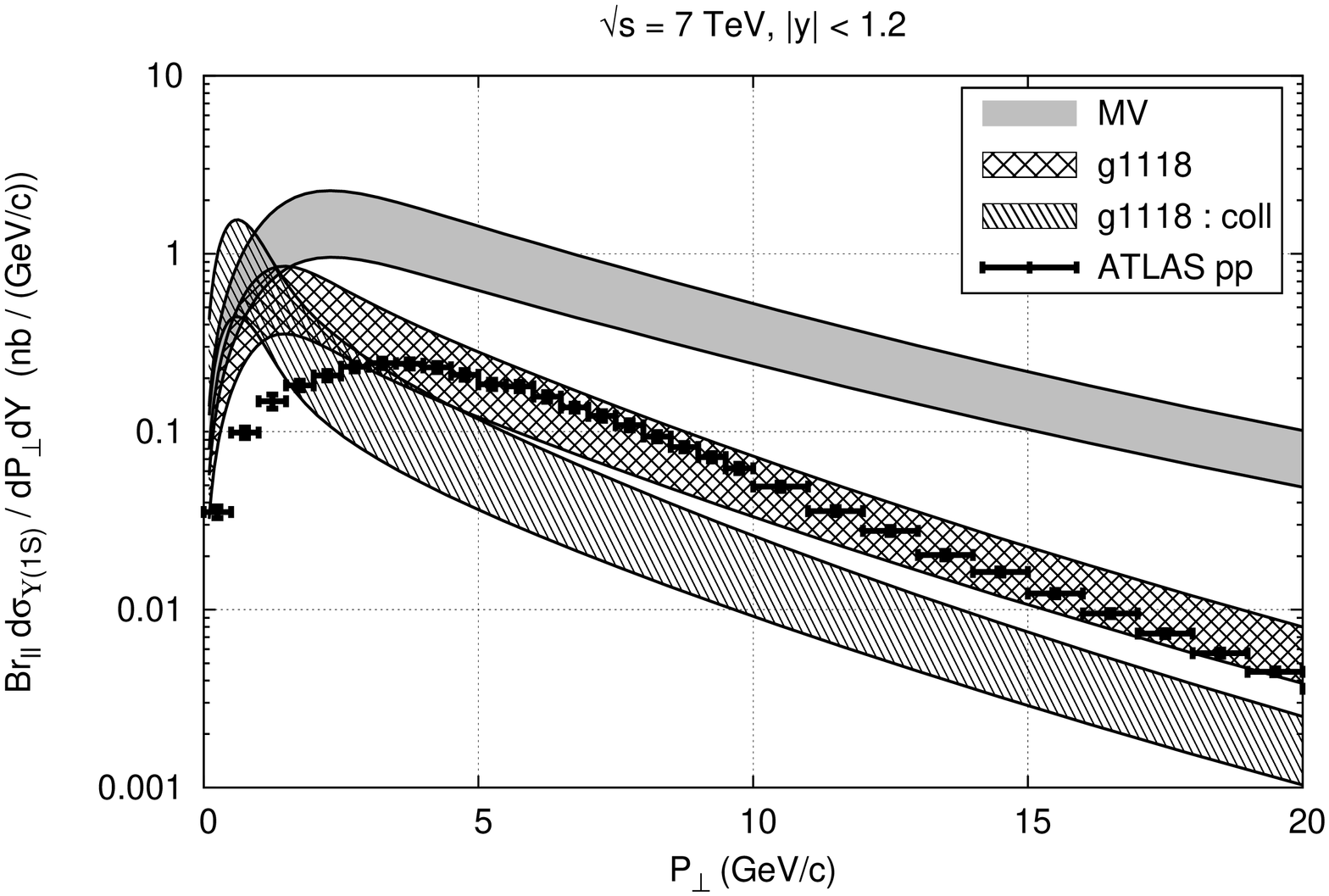}}
\resizebox*{!}{5.5cm}{\includegraphics[angle=270]{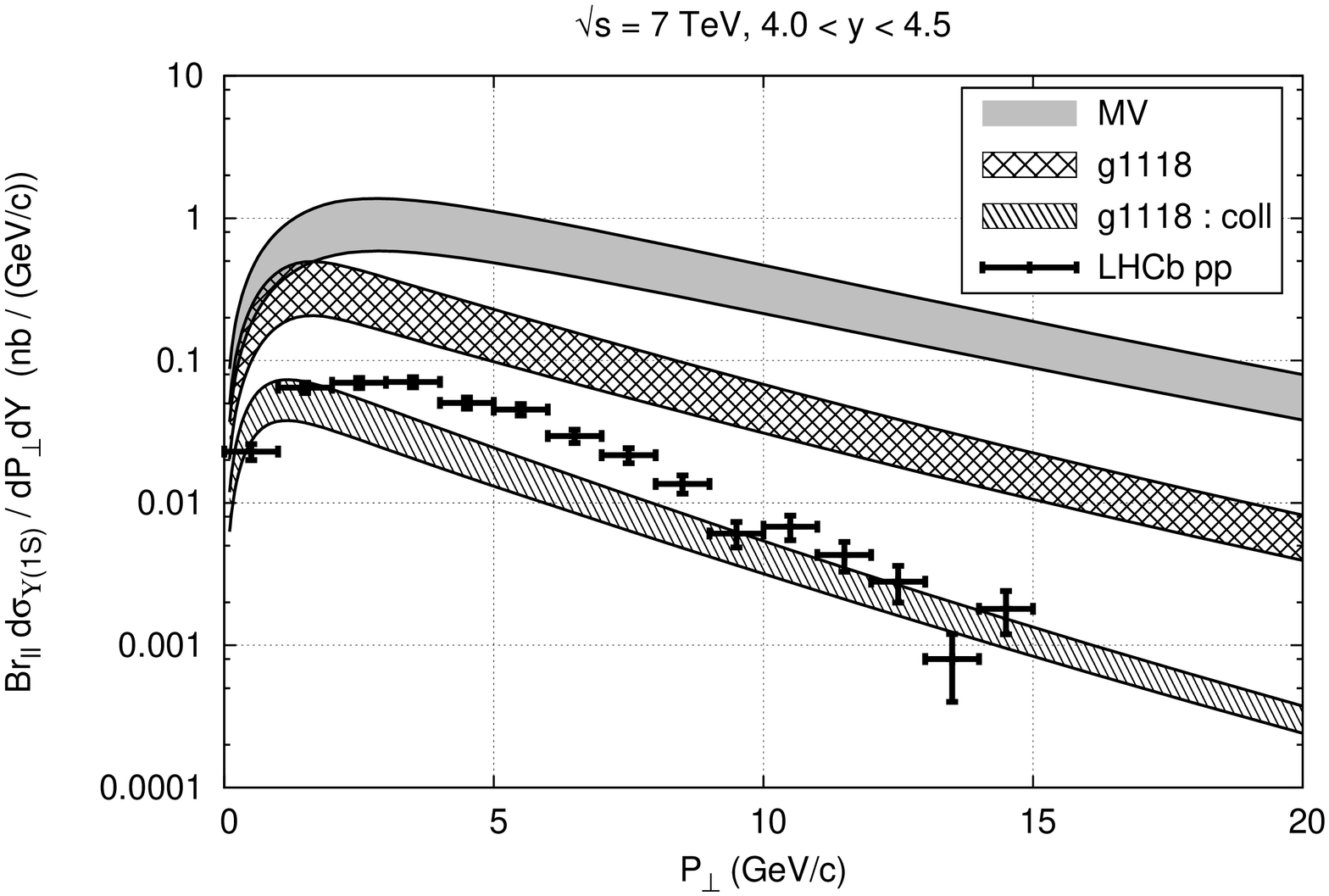}}
\end{center}
\caption{
Transverse momentum spectrum of $\Upsilon(1S)$ in di-lepton channel 
in pp collisions at $\sqrt{s}=7$ TeV for (a) $|y|<1.2$ and 
(b) $4<y<4.5$.
Notations are the same as in Fig.~\ref{fig:Jpsi_pt_rhic_pp}.
Data from \cite{Aad1,Aaij1}.}
\label{fig:Upsilon-pp}
\end{figure}

\begin{figure}[tbp]
\begin{center}
\resizebox*{!}{5.5cm}{\includegraphics[angle=270]{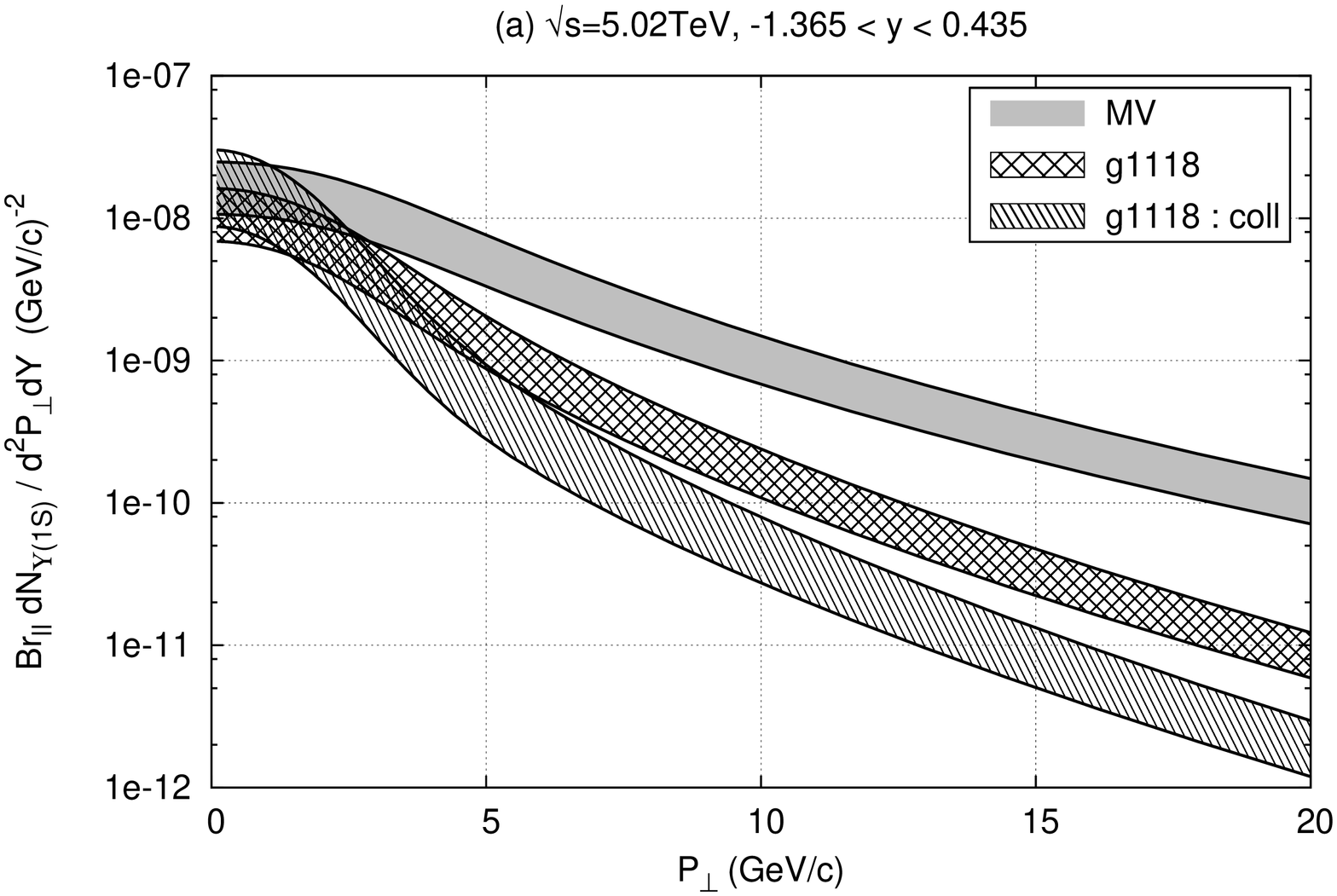}}
\resizebox*{!}{5.5cm}{\includegraphics[angle=270]{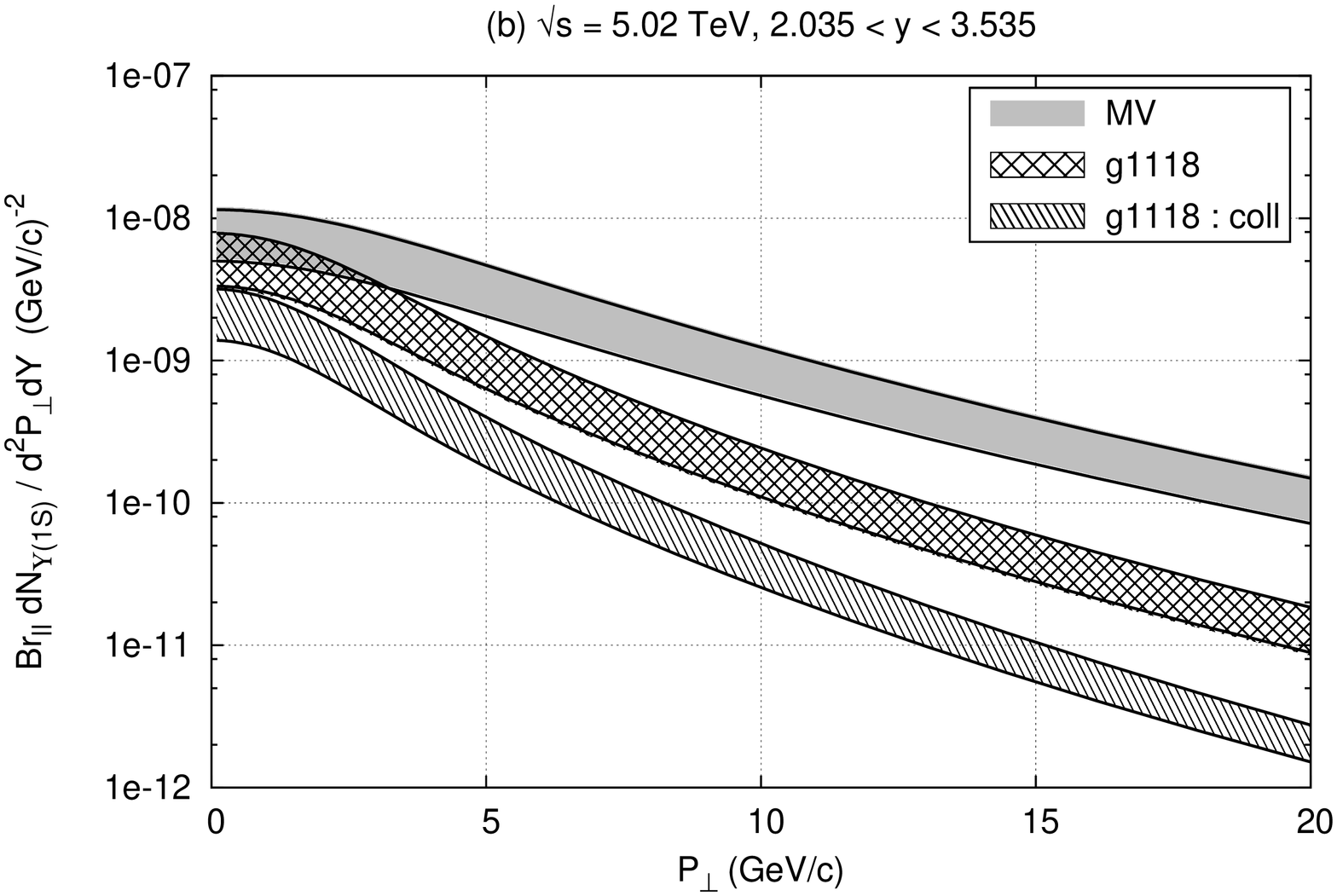}}
\end{center}
\caption{
Transverse momentum spectrum of $\Upsilon(1S)$ in di-lepton channel 
in pA collisions at $\sqrt{s}=5.02$ TeV 
for (a) $-1.4<y<0.4$ and (b) $2<y<3.5$. 
Notations are the same as in Fig.~\ref{fig:Jpsi_pt_rhic_pp}.
}
\label{fig:Upsilon-pt-lhc}
\end{figure}

Next we consider $\Upsilon(1S)$ production.
Non-linear effects are generally suppressed by the
inverse power of the heavy quark mass.
However, since the bottom quark mass $m_b$ 
is just three times as heavy as the charm quark mass $m_c$, 
the relevant value of $x$ for the $\Upsilon(1S)$ production
becomes larger by the same factor at low $P_\perp$, as compared
to the J/$\psi$. 
At the LHC energy, this $x$ value may be still small enough for 
multiple scatterings and saturation to be relevant in
the $\Upsilon$ production.

We plot the $P_\perp$ spectrum of $\Upsilon(1S)$ in pp and pA
collisions at $\sqrt{s}=7$ and 5.02 TeV,
respectively, in Figs.~\ref{fig:Upsilon-pp} and \ref{fig:Upsilon-pt-lhc}, 
together with the data measured by ATLAS and LHCb\cite{Aad1,Aaij1}
for the pp case.
Here we have chosen the CEM parameter as $F_{\Upsilon(1S)}=0.01$, 
and varied $m_b$ from 4.5 to 4.8 GeV.
Other notations are the same
as in the J/$\psi$ case.
In pp collisions, 
the coincidence between the model and the data for $\Upsilon(1S)$
state is not as good as that for J/$\psi$ at low $P_\perp$ and at
forward rapidity.

\begin{figure}[tbp]
\begin{center}
\resizebox*{!}{5.5cm}{\includegraphics[angle=270]{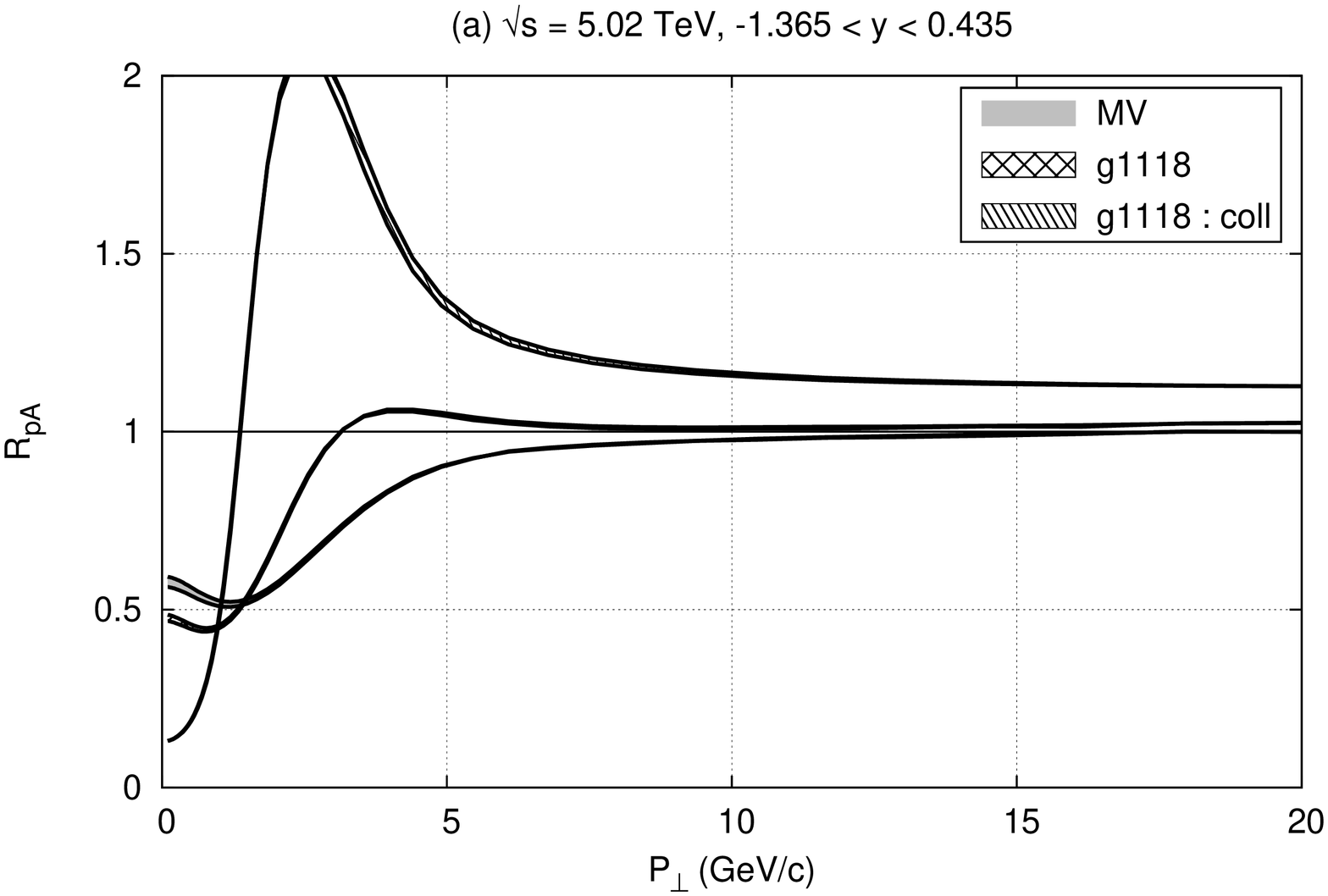}}
\resizebox*{!}{5.5cm}{\includegraphics[angle=270]{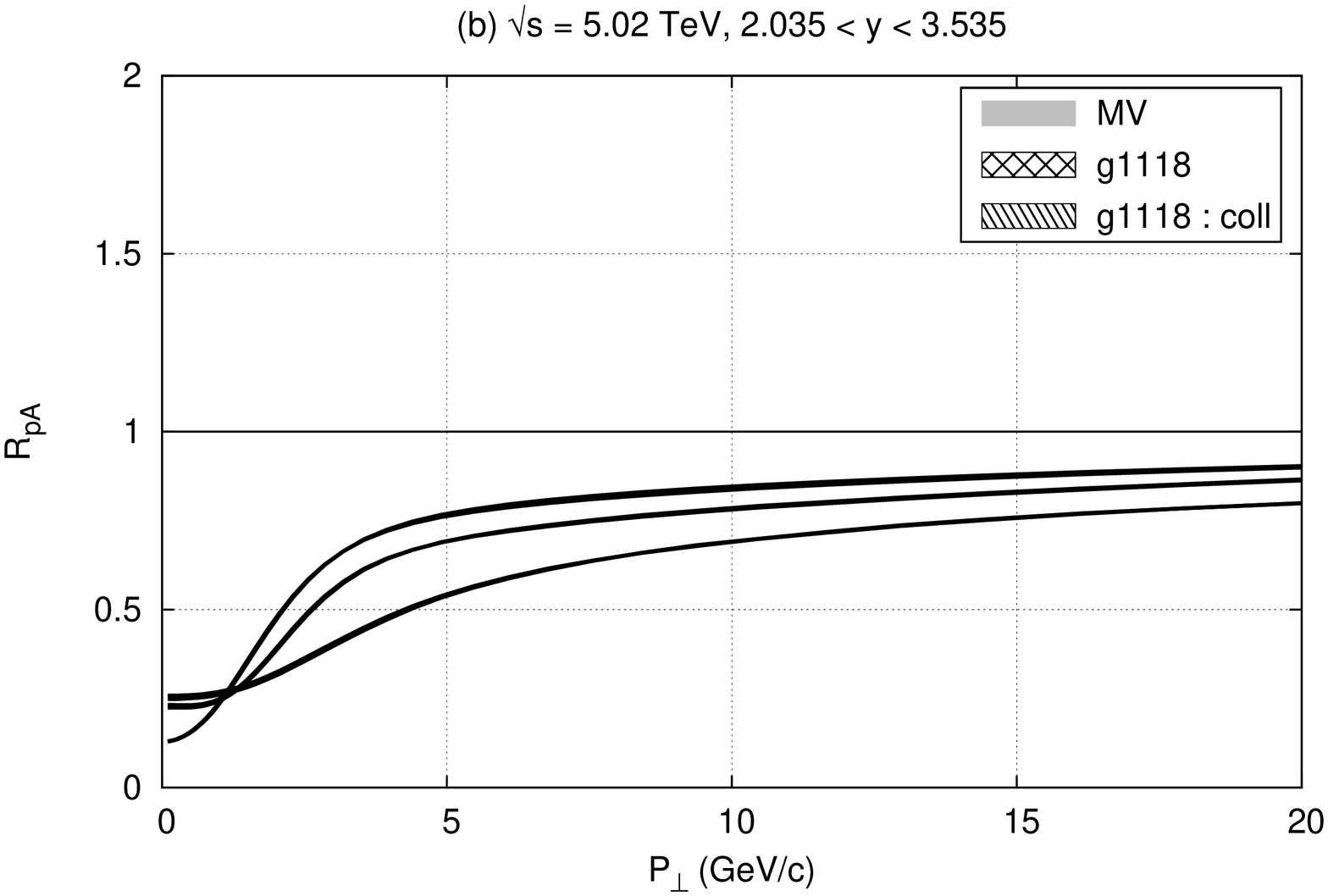}}
\end{center}
\caption{The ratio $R_\text{pA}$ for $\Upsilon(1S)$ 
at $\sqrt{s}=5.02$ TeV as a function of the transverse momentum.
(a) $-1.4<y<0.4$ and (b) $2<y<3.5$. 
Notations are the same as in Fig.~\ref{fig:Jpsi_pt_rhic_pp}.
}
\label{fig:RpA_Upsilon_lhc}
\end{figure}

We present in Fig.~\ref{fig:RpA_Upsilon_lhc}
the nuclear modification factor $R_\text{pA}$ for $\Upsilon(1S)$
as a function of $P_{\perp}$.
The model uncertainty from the quark mass value and the factorization
scale would cancel out by taking the ratio of the cross-sections
in the pp and pA collisions.
Indeed, each band collapses into a thin line whose width is almost
unnoticeable.

This result for $\Upsilon(1S)$ is qualitatively very similar to 
that for J/$\psi$.
At mid-rapidity, we see a suppression $R_\text{pA}$ in low $P_\perp$ region below
5 GeV, while it turns back to unity at larger $P_\perp$.
Only in the collinear approximation, we see the Cronin-like
enhancement, which is largely caused by the dip structure 
in the proton uGD at moderate $x_1$.
At forward rapidities $2<y<3.5$,
the $\Upsilon$ production is suppressed in a wide $P_\perp$
region from 0 to 20 GeV, irrespective of the model uGD's, g1118 or MV,
or of the use of collinear approximation.
In the forward region, $\Upsilon(1S)$ production has the
sensitivity to the small-$x$ evolution of uGD in the nucleus. 

We have also checked the initial-scale ($Q_{s0,A}^2$)
dependence of $R_\text{pA}$ for $\Upsilon(1S)$
by comparing the result with
$Q_{s0,A}^2=4Q_{s0,p}^2$ and $6Q_{s0,p}^2$ to find that the change is
very similar to the case with J/$\psi$ (Fig.~\ref{fig:RpA_Jpsi_lhc_46}).

\subsection{Rapidity dependence of $R_\text{pA}$ of J/$\psi$ and $\Upsilon$}

\begin{figure}[tbp]
\begin{center}
\resizebox*{!}{5.5cm}{\includegraphics[angle=270]{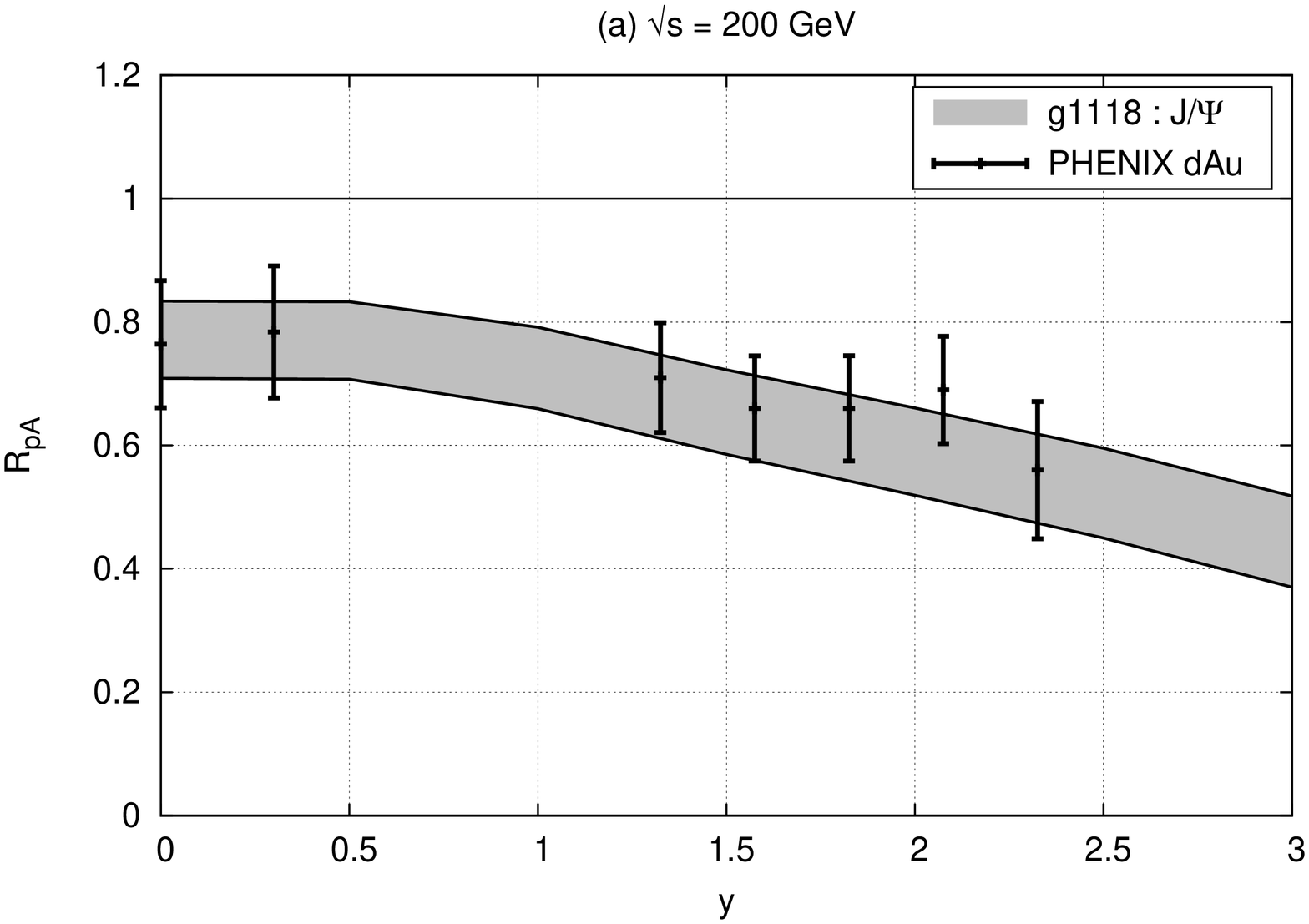}}
\resizebox*{!}{5.5cm}{\includegraphics[angle=270]{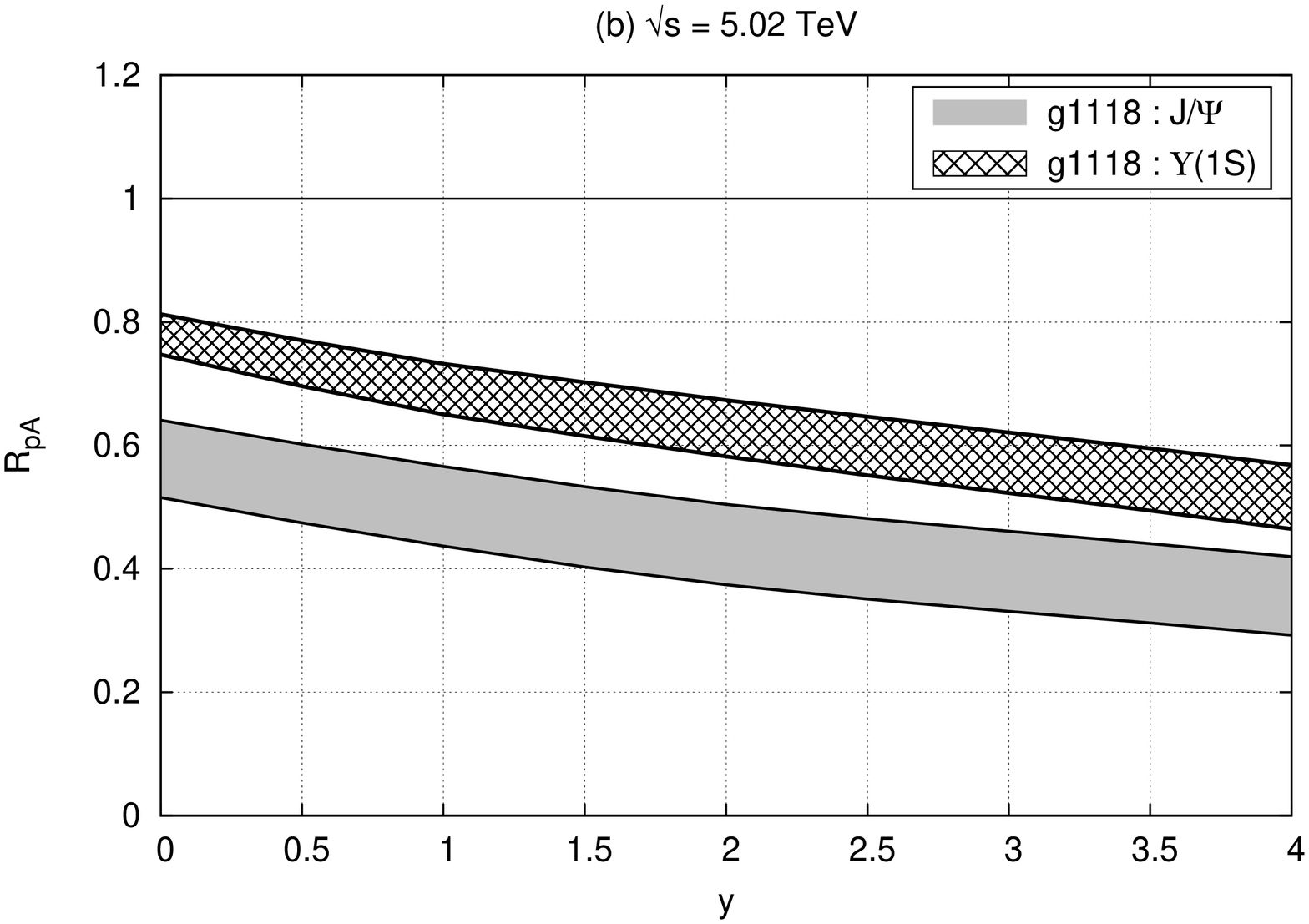}}
\end{center}
\caption{Nuclear modification factor $R_\text{pA}$ for J/$\psi$ 
integrated over $P_\perp$ as a function of rapidity, 
in pA collisions 
at (a) $\sqrt{s}=200$ GeV and (b) $\sqrt{s}=5.02$ TeV.
The band includes uncertainty for $m_c=1.2$ GeV to 1.5 GeV
and $Q_{s0,A}^2= (4-6) Q_{s0,p}^2$.
In (b), $R_\text{pA}$ of $\Upsilon(1S)$ is also shown 
where we vary $m_b=4.5$ to 4.8 GeV.
RHIC data from \cite{Adare4}. 
}
\label{fig:RpA_Jpsi_y_dep}
\end{figure}

We study the rapidity dependence of the ratio
$R_\text{pA}$ integrated over $P_\perp$.
The computation is performed with set g1118.
In Fig.~\ref{fig:RpA_Jpsi_y_dep} shown is
$R_\text{pA}(y)$ of J/$\psi$ at $\sqrt{s}=0.2$ and 5.02 TeV, 
together with that of $\Upsilon(1S)$ for the latter.
Note that our assumption of dilute-dense colliding system
applies only in the positive rapidity region ($y>0$), 
especially for pp,
which is needed in the denominator of $R_\text{pA}$.

We see systematically a stronger suppression of $R_\text{pA}$ 
as the rapidity increases both at RHIC and LHC energies.
This is in accord with $x$-evolution of uGD in the heavy target.
$R_\text{pA}(y)$ of J/$\psi$ flattens out at $y \lesssim 1$ at
RHIC energy because
the J/$\psi$ is produced there by the gluons with $x_2>x_0$
and we freeze the saturation scale to its initial value 
$Q_s^2(x>x0) = Q_{s,0}^2$.

Comparing the results of 
J/$\psi$ and $\Upsilon(1S)$ at LHC, 
we note that the suppression of $\Upsilon(1S)$ is 
smaller than that of J/$\psi$, but is still significant to be observed.
It would be quite important to study these systematics in experimental
data in order to quantify the saturation effects in the heavy nuclear target.

\subsection{$Q_{s0,A}^2$ dependence of $R_\text{pA}$}

\begin{figure}[tbp]
\begin{center}
\resizebox*{!}{5.5cm}{\includegraphics[angle=270]{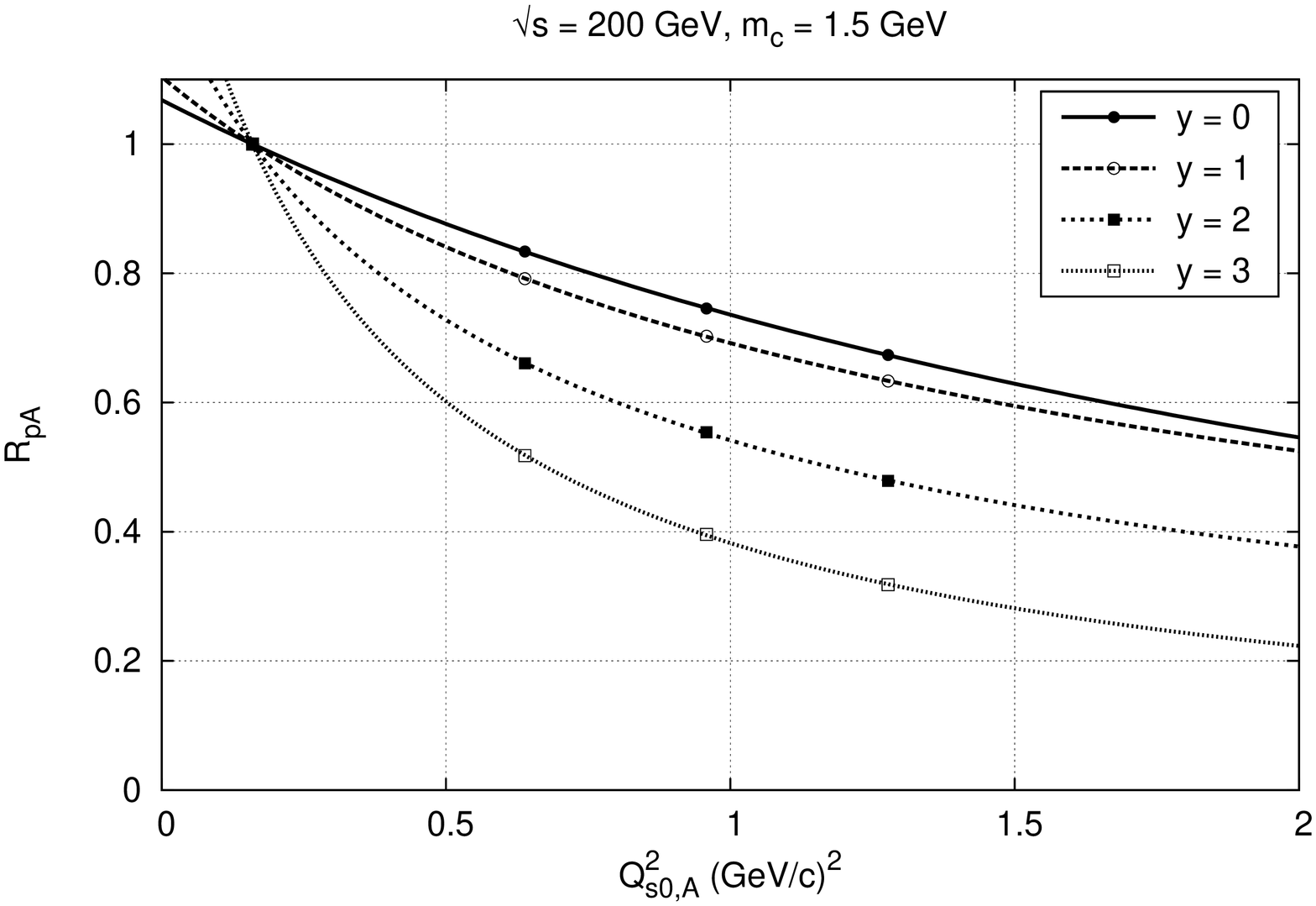}}
\resizebox*{!}{5.5cm}{\includegraphics[angle=270]{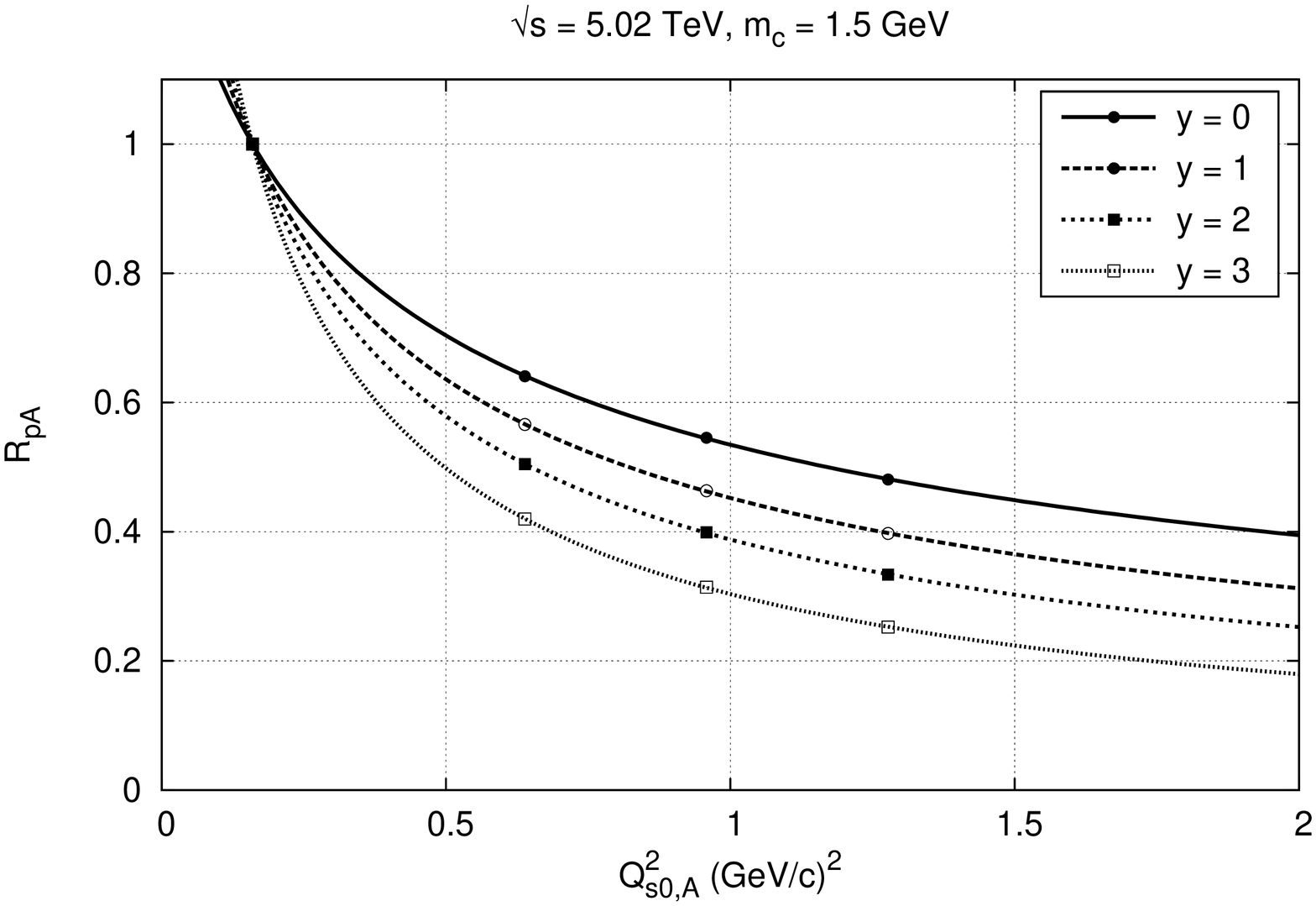}}
\end{center}
\caption{Nuclear modification factor $R_\text{pA}$ for J/$\psi$
as a function of $Q_{s0,A}^2$ at
$y=0,1,2$ and $3$ at $\sqrt s=200$ GeV (left) and $\sqrt s=5.02$ TeV
(right). Fitted curves are also shown.}
\label{fig:RpA_Qs_Jpsi}
\end{figure}

\begin{figure}[tbp]
\begin{center}
\resizebox*{!}{5.5cm}{\includegraphics[angle=270]{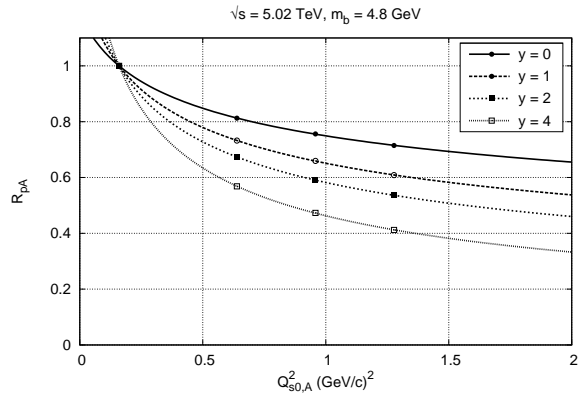}}
\end{center}
\caption{Nuclear modification factor $R_\text{pA}$ for $\Upsilon(1S)$
as a function of $Q_s^2$ at $y=0,1,2$ and $4$
at $\sqrt s=5.02$ TeV.
}
\label{fig:Qs-Upsilon}
\end{figure}

It would be interesting to study the dependence of 
$R_\text{pA}$ on the saturation scale parameter
$Q_{s0,A}^2$, which may be translated to
the effective thickness of the target.
We compute $R_\text{pA}$ of J/$\psi$ and $\Upsilon(1S)$
 integrated over $P_\perp$
as a function of $Q_{s0,A}^2$
at several values of $y$. 
We fix here the uGD set g1118  and 
the quark masses as $m_c=1.5$ GeV and $m_b=4.8$ GeV.
In Fig.~\ref{fig:RpA_Qs_Jpsi} we plot 
$R_\text{pA}$ of J/$\psi$ at $\sqrt{s}=0.2$ and 5.02 TeV.
We found that 
for each rapidity $Q_{s0,A}^2$-dependence of $R_\text{pA}$ can be 
fitted nicely by a model function:
\begin{align}
R_\text{pA}= \frac{a}{(b+Q_{s0,A}^2)^\alpha}
\label{eq:RpA-fit}
\end{align}
with $a$, $b$ and $\alpha$ being parameters.
This functional form is motivated by 
QCD analog of {\it superpenetration} of a electron-positron pair
through a medium\cite{Fujii,FujiiGV2}.
We show $\Upsilon(1S)$ result in Fig.~\ref{fig:Qs-Upsilon}
with fitted curves.
The stronger suppression at the larger value of $Q_{s0,A}^2$ is
naturally understood as a result of stronger multiple scatterings and 
saturation effects in the heavier target.

Energy and rapidity dependences may be qualitatively 
inferred through the increase of $Q_{s,A}^2(y)$ with increasing $y$.
Thus we tried to fit the rapidity dependence of $R_\text{pA}$ 
by replacing in Eq.~(\ref{eq:RpA-fit}) 
$Q_{s0,A}^2 \to Q_{s0,A}^2 e^{\lambda y}$ 
with a free parameter $\lambda$, 
but it was only unsatisfactory.
We remark here 
that quarkonium suppression due to parton saturation
in our treatment is twofold:
a relative depletion of the gluon source and 
multiple scatterings of the quark pair in the target.
The latter disturbs the boundstate formation,
by increasing the pair's invariant mass on average
in CEM\cite{Fujii3}.
It appears hard to describe energy and rapidity dependence of the
suppression at the same time through a single function $Q_{s,A}^2(x)$.

\subsection{$P_\perp$ broadening}

\begin{figure}[tbp]
\begin{center}
\resizebox*{!}{5.5cm}{\includegraphics[angle=270]{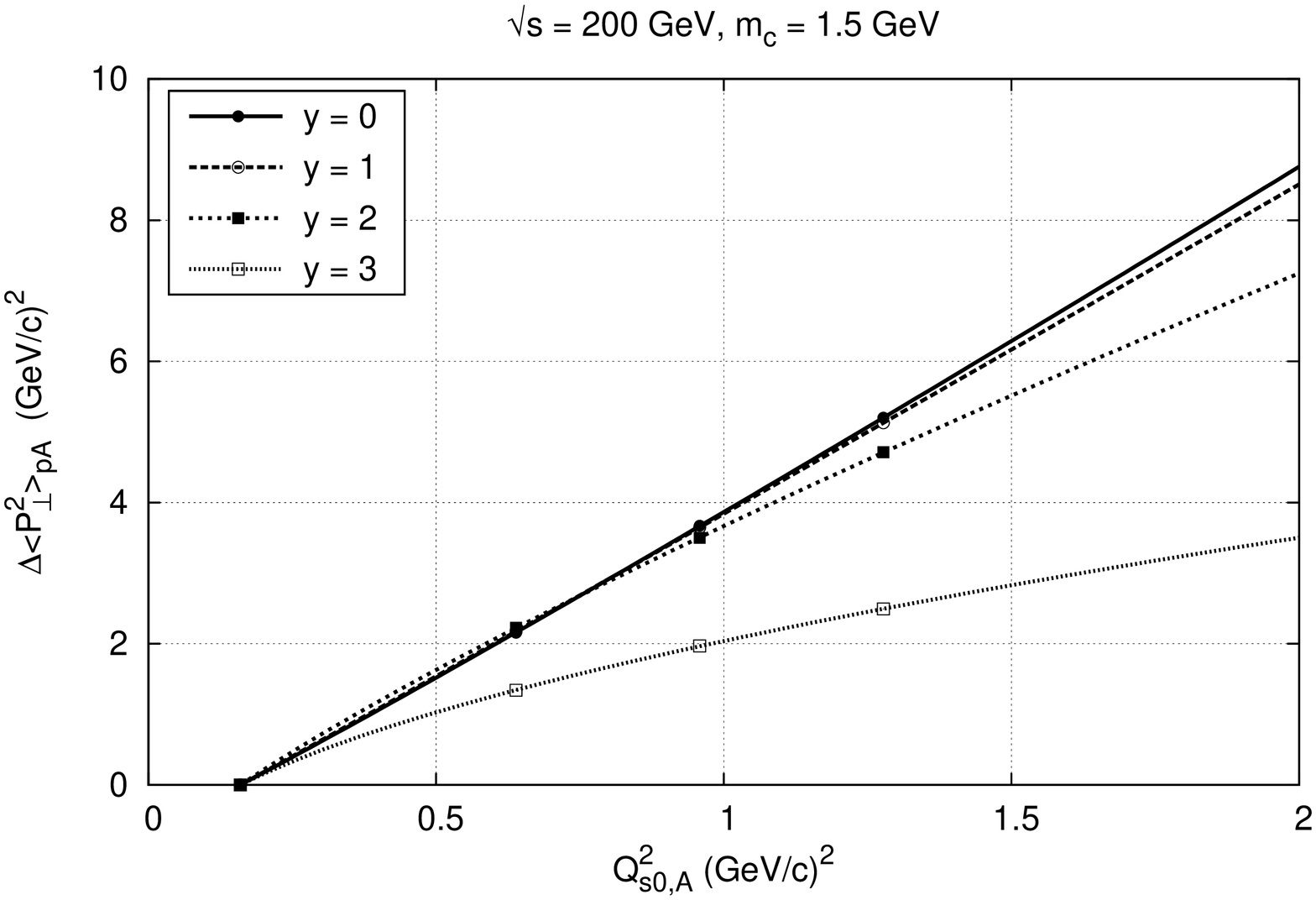}}
\resizebox*{!}{5.5cm}{\includegraphics[angle=270]{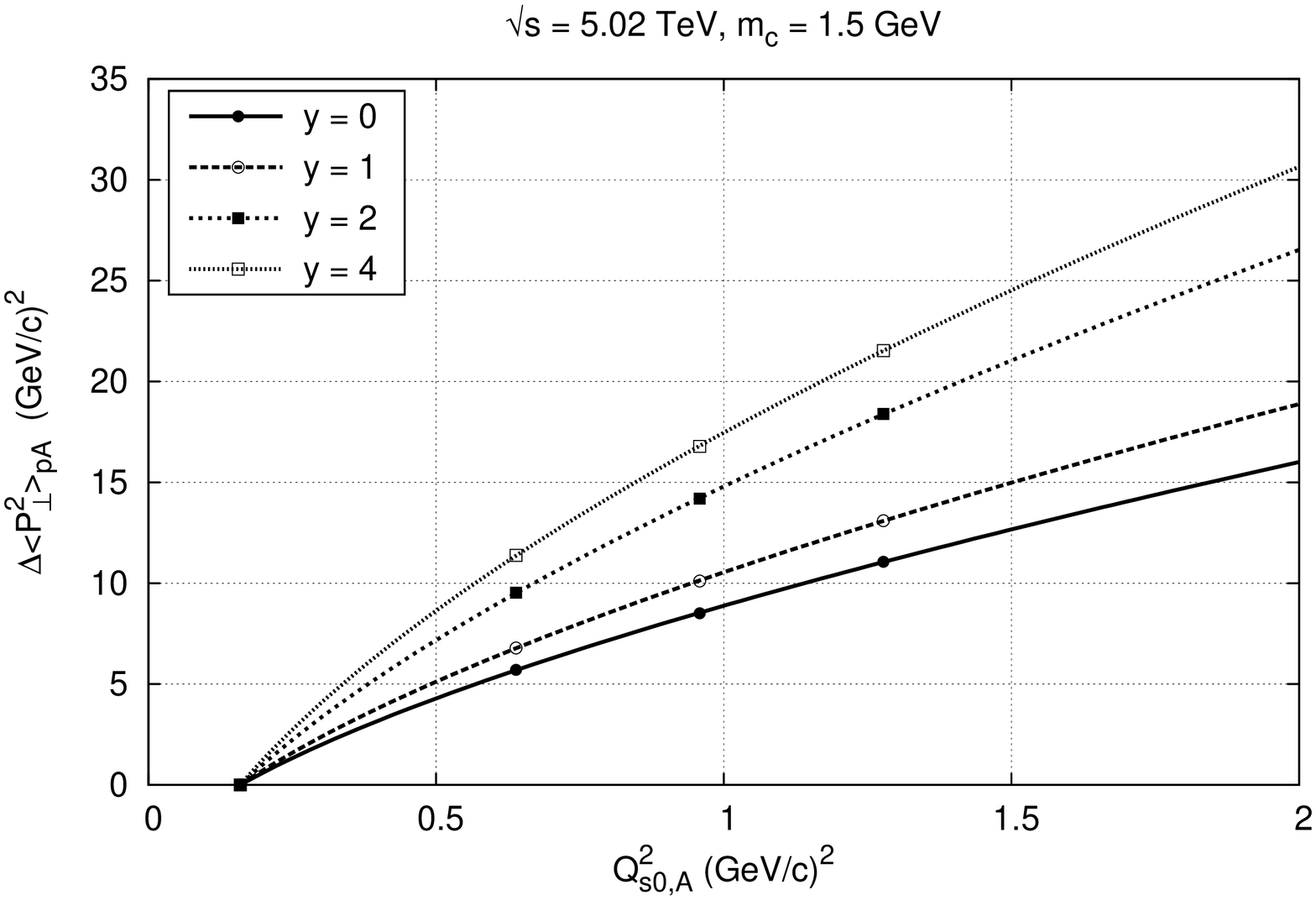}}
\end{center}
\caption{Mean transverse momentum square
$\Delta \langle P^2_\perp\rangle_\text{pA}$
of J/$\psi$ as a function of $Q_{s0,A}^2$ at $\sqrt s=200$ GeV (left) 
and $\sqrt s=5.02$ TeV (right).
Fit with a form $a [(Q_{s0,A}^2/Q_{s0,p}^2)^\alpha -1 ]$ is also shown.}
\label{fig:Pt_Qs_Jpsi}
\end{figure}

Finally, 
we study the mean transverse momentum of quarkonium
in pA collisions.
The momentum broadening in the nuclear target has been 
discussed in the literature\cite{Brambilla:2004wf}.
In our framework,
the multiple scatterings of the incident gluon and
the produced quark pair in the nuclear target,
encoded in $U$ and $\tilde U$ terms in Eq.~(\ref{eq:Mf-final-1})
respectively, cause the momentum broadening of the pair.
Typical momentum transfer of the multiple scatterings
in the nucleus should be characterized
by the saturation scale $Q_{s,A}(x_2)$.
We define here the broadening of $P_\perp$
as the deviation of the mean transverse momentum squared 
$\langle P_\perp^2 \rangle$
of J/$\psi$ in pA collisions 
from that in pp collisions:
\begin{eqnarray}
\Delta \langle P^2_\perp\rangle_\text{pA} \equiv
\langle P^2_\perp\rangle_\text{pA}
-\langle P^2_\perp\rangle_\text{pp}
=
 \frac{\int d\sigma_\text{pA} P_\perp^2 }{\int  d\sigma_\text{pA} }
-\frac{\int d\sigma_\text{pp} P_\perp^2 }{\int  d\sigma_\text{pp} }. 
\label{eq:Pt_broadening}
\end{eqnarray}

In Fig.~\ref{fig:Pt_Qs_Jpsi}
we plot $\Delta \langle P_\perp^2 \rangle_\text{pA}$ 
as a function of $Q_{s0,A}^2$.
We use uGD set g1118
with the quark masses $m_c=1.5$ GeV and $m_b=4.8$ GeV.
We have found that for each rapidity
the $Q_{s0,A}^2$ dependence of 
the broadening can be fitted 
in a simple form:
\begin{align}
\Delta\langle P^2_\perp\rangle_\text{pA}
=a[(Q_{s0,A}^2/Q_{s0,p}^2)^\alpha -1]
\end{align}
with $a$ and $\alpha$ being parameters.

At $\sqrt{s}=200$ GeV, 
the broadening at mid-rapidity 
is obviously linear in $Q_{s0,A}^2$, 
which indicates the random walk nature of the multiple scatterings
in the momentum space.
In the forward region, we naively expected an increase 
of the mean momentum by the stronger scatterings, 
but actually found the opposite, 
i.e., a decrease from the mid-rapidity value.
We interpret this as the effect of kinematical boundary of $x_1$
in the forward region (see Fig.~\ref{fig:x2-coverage}).

The measured value of $\Delta \langle P_\perp^2 \rangle_\text{dA}$
at RHIC\cite{Adare3} seems to be smaller by a factor of 5 than that
in Fig.~\ref{fig:Pt_Qs_Jpsi},
if we naively translate $Q_{s0,A}^2$ to the centrality parameter
$N_\text{coll}$ evaluated for dAu collisions.
This strong broadening originates probably from the fact that
our model has too hard $P_\perp$ spectrum at RHIC energy.
But it is at least consistent with data that
$P_\perp$ broadening at forward rapidities $\sim 2$ is weaker
than that at mid-rapidity $y \sim 0$.

\begin{figure}[tbp]
\begin{center}
\resizebox*{!}{5.5cm}{\includegraphics[angle=270]{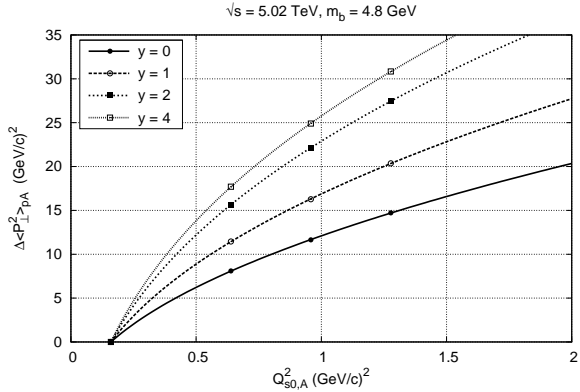}}
\end{center}
\caption{
Mean transverse momentum square
$\Delta \langle P^2_\perp\rangle_\text{pA}$ of
$\Upsilon(1S)$  
as a function of $Q_s^2$ at $\sqrt s=5.02$TeV.
Fit with a form $a [(Q_{s0,A}^2/Q_{s0,p}^2)^\alpha -1 ]$ is also
shown.}
\label{fig:Upsilon-Pt-Qs}
\end{figure}

At $\sqrt{s}=5.02$ TeV, a wider phase space opens up and we instead see
an increase of the mean momentum of J/$\psi$ as moving to the forward
rapidity region.  
We have checked that $\Delta \langle P_\perp^2 \rangle_\text{pA}$
gets back to be smaller at $y=6$ than 
that of mid-rapidity, just as seen in the case of $\sqrt{s}=200$ GeV. 
Nonlinear dependence on $Q_{s0,A}^2$ may imply the different
evolution speed of multiple-scattering strength
for different initial values $Q_{s0,A}^2$. 
The result for $\Upsilon(1S)$ 
in Fig.~\ref{fig:Upsilon-Pt-Qs} is similar to the J/$\psi$ case, but 
interestingly the broadening becomes more remarkable;
The heavier bottom quark pair can acquire the larger transverse momentum
$P_\perp$ in multiple scatterings before going beyond the threshold set on
the pair's invariant mass $M^2 < 4 M_B^2$

\section{Conclusion and Outlook}

Quarkonium production in proton-lead collisions
provides us with a good opportunity to study 
the saturation phenomenon in the incident nucleus
thanks to the wide kinetic reach at the LHC.
We have computed the J/$\psi$ and $\Upsilon(1S)$
production in pA collision
at collider energies within CEM based on the CGC quark pair production,
and have discussed sensitivity of the
quarkonium observables to the parton saturation in the target nucleus.
We have presented the calculations with the uGD set g1118
which is constrained with DIS data at $x<x_0=0.01$,
with and without the collinear approximation, 
and the calculations with the model uGD set MV for comparison.

At the RHIC energy $\sqrt{s}=200$ GeV, 
the J/$\psi$ at mid-rapidity is produced not 
from the small-$x$ gluons, but from the moderate-$x$ gluons,
and  the $P_\perp$ spectrum in pp collisions 
is unfortunately sensitive to a unphysical dip 
structure of the uGD set g1118, which was constrained only
for $x<x_0$. 
We need better extrapolation of our framework to $x \gtrsim x_0$.
In pA collisions, multiple scatterings smear out the dip of the uGD
and the $P_\perp$ spectrum of J/$\psi$ becomes closer to the observed one
in dAu collisions.

At the LHC energy $\sqrt{s}=5.02$ TeV, the small-$x$ gluons dominate
the charm production, and we have found that 
our model with the uGD set g1118 works
for J/$\psi$ production in pp collisions both at 
mid- and forward-rapidities. Then we have shown our model prediction  
on J/$\psi$ production in pA collisions.
The ratio $R_\text{pA}(P_\perp)$ for J/$\psi$
shows a suppression for $P_\perp \lesssim 5$ GeV at
mid-rapidity due to saturation effects, 
and it is further suppressed in wider range of $P_\perp$
as moving to forward rapidities.

We have also shown that the $\Upsilon(1S)$ production in pA collisions
at the LHC has a good sensitivity to the gluon saturation of the nucleus,
provided that the effect is smaller than that in the J/$\psi$ case.
In our model, when integrated over $P_\perp$, 
the ratio $R_\text{pA}(y)$ for $\Upsilon$ at the LHC shows a
suppression similar to that of J/$\psi$ at RHIC energy.

Transverse momentum broadening 
$\Delta \langle P_\perp^2 \rangle$ of the quarkonium
shows an increasing behavior as a function of $Q_{s0,A}^2$.
Because our model gives harder $P_\perp$ spectrum than the data, 
the broadening is likely to be overestimated at RHIC energy 
in our calculation.
However, it is still
interesting to notice that at RHIC energy the broadening
becomes weaker at forward rapidity than at mid-rapidity
due to the kinematical constraint.
At LHC energy, on the other hand,
we expect the increase of the broadening at
forward rapidities because
of the larger saturation scale $Q_s^2(x_2)$ and 
wider kinematic coverage of the LHC.
Transverse momentum broadening is also investigated recently 
by taking account of the multiple scatterings in the target 
in \cite{Kang:2012am}.

In conclusion,
we have numerically studied the quarkonium production in pA collisions
at the RHIC and LHC, within CEM based on the CGC quark-pair production 
energies, 
and have quantifies the effects of saturation and multiple scatterings
in the target nucleus on the J/$\psi$ and $\Upsilon$ observables.
Comparison of our results with experimental data at the LHC
must be very important to access the relevance of the saturation
physics in the quarkonium production.

In this work we have employed CEM
to describe the nonperturbative formation of the quarkonia. 
In fact, quarkonium formation is one of the challenges in QCD, even in
pp collisions, and CEM replaces this just by a probability constant
$F_{{\rm J}/\psi}$. 
Non-relativistic QCD framework has recently been extended to NLO,
which improves our understanding of quarkonium production 
significantly\cite{Ma,Buten}.  
As a first step in this direction we plan to
 match the quark-pair production from CGC  
onto the non-relativistic QCD approach.

We can investigate also the production of open heavy flavor mesons 
in our framework.
Modification of $P_\perp$ spectrum and $D\bar D$
correlations will also provide very useful information on the
saturation in the target nucleus and a good benchmark for the
energy loss and collective flow measurements of $D$ and $B$ mesons.
We will report this elsewhere\cite{FujiiW2}.

\section*{Acknowledgments}
We are very grateful to 
J.~Albacete, A.~Dumitru, F.~Gelis, K.~Itakura, Y.~Nara, R.~Venugopalan
for useful discussions and collaborations on related topics.
This work was partially supported by Grant-in-Aids 
for Scientific Research ((C) 24540255) of MEXT.

\end{document}